\documentclass[pdflatex,sn-mathphys-num,iicol]{sn-jnl}% Math and Physical Sciences Numbered Reference Style
\usepackage{url}
\usepackage{graphicx}%
\usepackage{multirow}%
\usepackage{amsmath,amssymb,amsfonts}%
\usepackage{amsthm}%
\usepackage{mathrsfs}%
\usepackage[title]{appendix}%
\usepackage{xcolor}%
\usepackage{textcomp}%
\usepackage{manyfoot}%
\usepackage{booktabs}%
\usepackage{url}
\usepackage{braket}
\usepackage{mathrsfs}
\usepackage{algpseudocode}%
\usepackage{listings}%

\usepackage[table]{xcolor} 
\usepackage{tabularx}
\UseRawInputEncoding
\usepackage[ruled,vlined,linesnumbered]{algorithm2e}

\theoremstyle{thmstyleone}%
\theoremstyle{thmstyletwo}%

\theoremstyle{thmstylethree}%

\begin{document}

\title{Reconfigurable Bus-based Quantum Router for Modular Superconducting Processors}

\author[1]{\fnm{Benzheng} \sur{Yuan}}
\author[1]{\fnm{Chaojie} \sur{Zhang}}
\author[1]{\fnm{Yangyang} \sur{Fei}}

\author[1]{\fnm{Chuanbing} \sur{Han}}

\author[1]{\fnm{Haoran} \sur{He}}

\author[1]{\fnm{Huihui} \sur{Sun}}

\author[1]{\fnm{Bo} \sur{Zhao}}
\author[1]{\fnm{Fudong} \sur{Liu}}
\author*[1]{\fnm{Weilong} \sur{Wang}}\email{wangwl19888@163.com}
\author*[1]{\fnm{Zheng} \sur{Shan}}\email{shanzhengzz@163.com}

\affil[1]{\orgdiv{Laboratory for Advanced Computing and Intelligence Engineering}, \orgname{Information Engineering University}, \orgaddress{\street{No. 62 Science Avenue}, \city{Zhengzhou}, \postcode{450001}, \state{Henan}, \country{China}}}

\abstract{
Scaling superconducting quantum processors requires interconnects that provide both non-local connectivity and parallel entangling operations. While current quantum routers offer greater connectivity than nearest-neighbour topologies, their support for parallel, independently addressable two-qubit gates remains fundamentally limited, imposing severe compilation overheads that constitute a critical bottleneck. Here we introduce a bus-based reconfigurable quantum router that enables parallel controlled-$Z$ (CZ) gates for modular superconducting processors. Constructed from a flux-tunable SQUID network, the router selectively connects interface qubits to two shared buses, allowing destructive interference to suppress idle interactions while supporting two disjoint CZ gates in parallel. Full-system Hamiltonian simulations yield parallel-gate infidelities ranging from $7.9\times10^{-4}$ to $1.8\times10^{-3}$ under the operating conditions considered, and an open-system analysis identifies the coherence requirements for high-fidelity operation. We further assess the circuit-level consequences using hardware-aware compilation and resource-constrained scheduling. For the 36-qubit quantum Fourier transform (QFT), QAOA-MaxCut, and random-pairing circuits, the router reduces the median SWAP count by up to $34\%$ and the native CZ count by up to $20\%$ relative to a matched two-dimensional grid. Circuit-depth reductions are workload-dependent, reaching $20\%$ for QAOA-MaxCut but remaining negligible for the QFT despite its lower gate count. These results show that enhanced connectivity and schedulable parallelism provide complementary benefits for connectivity-intensive algorithms, establishing the router as a compiler-visible hardware resource and providing a scalable architectural pathway toward constructing highly connected modular superconducting quantum processors.
}

\keywords{Quantum Networks, Quantum Computation, Superconducting qubit}

\maketitle
\section{Introduction}\label{sec1}
The realization of fault-tolerant quantum computing (FTQC) is essential for achieving universal quantum computation \cite{williamson2026low,shor,krinner2022realizing,google2025quantum,Zhou2025,Acharya2023,whhNc}. Superconducting processors have recently achieved remarkable strides in qubit coherence and local gate fidelities \cite{siddiqi2021engineering, dd96-gcb6,ganjam2024surpassing,Gao2025,Tuokkola2025, Bland2025, Dane2026, Smith2020, Li2023,Li2019,PhysRevX.11.021058, rf7g-md41,lvb9-pfr3,PhysRevLett.125.240503, PhysRevLett.129.060501, Kim2022}. Concurrently, advances in fabrication have successfully scaled single-chip systems into the hundred-qubit regime and beyond. \cite{arute2019quantum,PhysRevLett.127.180501, Kim2023, PhysRevLett.134.090601, Niu2023, Gold2021}. However, constrained local connectivity has increasingly emerged as a fundamental bottleneck limiting overall processor performance. Mediating interactions between non-local qubits typically mandates extensive SWAP sequences, which inexorably inflates circuit depth, compounds control overhead, and accelerates error accumulation \cite{Vezvaee2026,CarreraVazquez2024,Steinberg2024, cswp-xy7k}. A pivotal challenge for scalable superconducting architectures is therefore to engineer selective, low-crosstalk, and parallelism-compatible non-local connectivity while strictly preserving qubit locality and individual device addressability \cite{PRXQuantum.4.010313, PhysRevLett.134.020801, 82cj-lfzy, Eickbusch2025,doi:10.1126/science.adz8659}.

Modular quantum routers offer a compelling paradigm to overcome these topological constraints: by synthesizing on-demand interactions across distributed modules, they substantially enhance hardware connectivity and gate flexibility without compromising modular scalability \cite{Zhou2023, PhysRevX.14.041030, wu2026efficientnqubitentanglingoperations}. Significant experimental progress has been achieved along this avenue within superconducting platforms. On the one hand, parametrically driven nonlinearities based on superconducting nonlinear asymmetric inductive elements (SNAILs) enable three-wave-mixing state routers that provide reconfigurable all-to-all photon routing among detachable quantum modules, thereby facilitating inter-module iSWAP operations \cite{Zhou2023,c661-yr2z,PhysRevX.14.031040,10.1063/1.4984142}. On the other hand, on-chip routers utilizing SQUID-based flux-tunable couplers have realized dynamically reconfigurable all-to-all coupling in modular processors \cite{PhysRevX.14.041030}, successfully demonstrating arbitrary-pair CZ and iSWAP gates alongside the generation of distributed multi-qubit entanglement \cite{wu2026efficientnqubitentanglingoperations}.

Existing routers predominantly focus on reconfigurable connectivity, quantum state transfer, and the serial execution of two-qubit gates, thereby falling short of fully unlocking the gate parallelism required by quantum algorithms. For practical quantum algorithms, such as QAOA-MaxCut and random-pairing workloads, the parallel execution of two-qubit gates is essential for compressing circuit depth \cite{PhysRevLett.133.150602, doi:10.1126/sciadv.adm6761}. Therefore, the core challenge for quantum routers lies in advancing from a mere “reconfigurable connector” to a functional processing unit capable of supporting parallel two-qubit operations.

In this work, inspired by classical routers, we propose a bus-based reconfigurable quantum router architecture that supports parallel CZ gates. A flux-tunable SQUID network selectively connects interface qubits to two shared buses, so that destructive interference suppresses residual idle interactions while two disjoint CZ gates can be executed in parallel. Taking the CZ gate as a representative example, we numerically simulate the system Hamiltonian to evaluate the parallel-gate control protocol and parallel gate fidelity. Our results demonstrate that this architecture can achieve parallel two-qubit operations with gate infidelities on the order of $10^{-3}$ within experimentally accessible parameter regimes. Open-system simulations further identify the coherence requirements for high-fidelity operation.

We then analyze whether these physical capabilities translate into algorithm-level gains. Using matched modular two-dimensional grid and router topologies, identical hardware-aware SABRE compilation settings, and resource-constrained as-soon-as-possible scheduling, we benchmark the 36-qubit QFT, QAOA-MaxCut, and eight-layer random-pairing circuits over 100 random compilation seeds \cite{10.1145/3297858.3304023}. Relative to the matched grid, the router reduces the median SWAP count by up to $34\%$ and the native CZ count by up to $20\%$ across all three benchmarks. The corresponding circuit-depth reduction is workload-dependent: QAOA-MaxCut and random-pairing circuits achieve median end-to-end reductions of $20\%$ and $15\%$, respectively, whereas the QFT shows no median depth reduction at this scale despite requiring fewer gates. These results show that enhanced connectivity and schedulable parallelism provide complementary benefits for connectivity-intensive algorithms. They establish the router as a compiler-visible hardware resource and provide a scalable architectural pathway toward highly connected modular superconducting quantum processors.

\section{Results}\label{sec2}
\subsection{Abstract model of router}
\label{Abstract model}

\begin{figure*}
\centering
\includegraphics[width=1\linewidth]{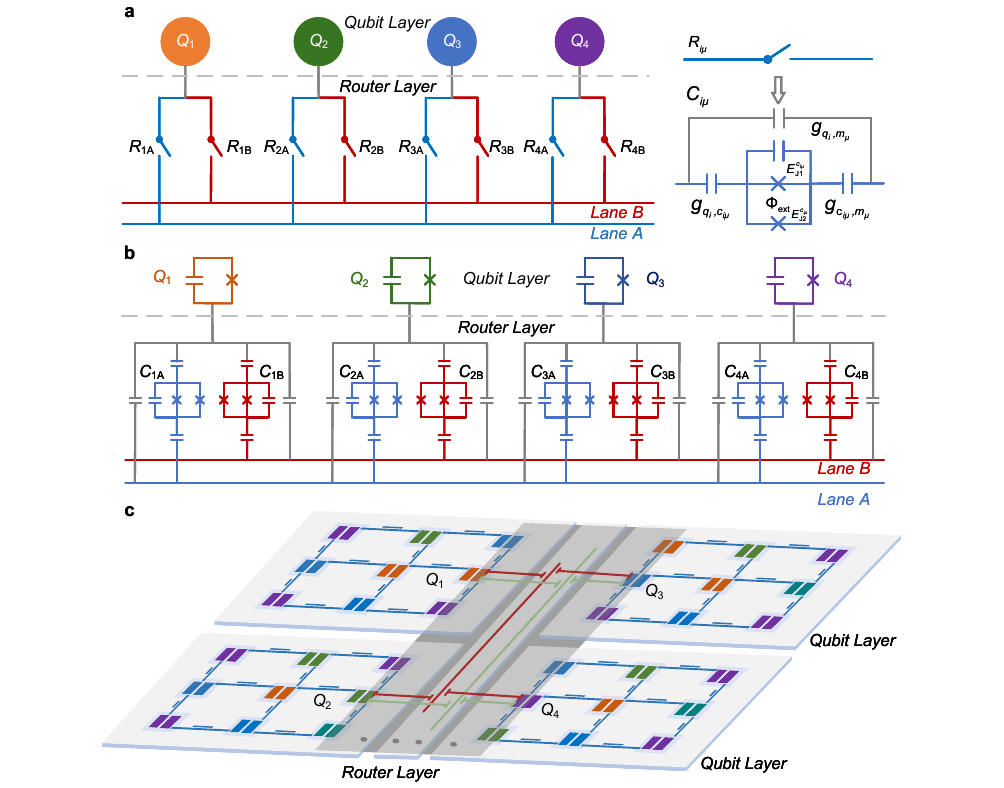}
\caption{\textbf{Conceptual schematic, equivalent circuit diagram, and flip-chip implementation blueprint of the quantum router.} 
\textbf{a,} Schematic of the quantum router processor. Four computational interface qubits ($Q_1$--$Q_4$) are connected to two transmission buses (Lane A and Lane B) through a switch array ($R_{i\mu}$, where $i \in \{1, 2, 3, 4\}$ denotes the qubit index and $\mu \in \{\mathrm{A}, \mathrm{B}\}$ denotes the routing bus channel). By selectively closing specific switches, designated qubit pairs can be coupled to the same bus. The inset on the right illustrates the physical implementation of the switch, in which the ideal switch is realized by a nonlinear tunable inductive element based on a superconducting quantum interference device (SQUID). 
\textbf{b,} Equivalent circuit model of the quantum router. The system consists of an upper qubit layer (Qubit Layer) and a lower routing layer (Router Layer). Four interface qubits ($Q_1$ to $Q_4$, color-coded) are capacitively coupled to two shared bus resonators (Lane A, blue; Lane B, red). The qubits and buses are interconnected through nonlinear tunable coupling networks ($C_{i\mathrm{A}}, C_{i\mathrm{B}}$) incorporating SQUIDs. By tuning the external magnetic flux to induce destructive interference among multiple coupling pathways, the effective coupling strength can be switched on and off. 
\textbf{c,} Flip-chip layout of the modular superconducting quantum processor. Four peripheral computational modules constitute the Qubit Layer, each containing a local lattice of nearest-neighbour-coupled transmon qubits dedicated to executing intra-module single- and two-qubit quantum logic operations. Each module is configured with an interface qubit ($Q_i$) connected to the central router layer (gray shaded area) via a multi-layer superconducting flip-chip process. The central router layer (gray shaded area) is integrated using superconducting flip-chip technology: an upper chip containing the array of SQUID coupling elements is flip-chip bonded onto the bottom substrate hosting the transmission buses, thereby establishing low-crosstalk interconnect channels between arbitrary non-nearest-neighbour modules.}
\label{fig1}

\end{figure*}

In classical computer networks, the primary function of a router is to enable efficient data exchange across multiple nodes \cite{11267428}. A typical classical router architecture relies on a shared-bus transmission mechanism: data enter through a designated input port, propagate along an internal bus, and are delivered to the target output port \cite{11267428, 52203}. To prevent data collisions and improve network throughput, such routers typically incorporate shared memory as a data buffer. Combined with multiple internal buses, this design enables parallel data transmission among multiple nodes.

Following this systems-engineering perspective, a bus-based quantum router can likewise be introduced to support entangling operations between qubits in different computational modules as modular superconducting quantum processors are scaled up. However, quantum routing is subject to constraints that have no classical analogue. In particular, the no-cloning theorem forbids copying an unknown quantum state, so a quantum router cannot rely on conventional store-and-forward mechanisms that duplicate or broadcast quantum data through shared classical memory. Although quantum memories can in principle buffer a quantum state without cloning it, practical superconducting routers are severely limited by finite coherence times and by the lack of scalable multimode quantum memories. In this work, we therefore adopt a bufferless, real-time control strategy that synthesizes low-crosstalk interaction channels between target modules on demand.

Inspired by classical multi-bus routing, we propose a bus-based quantum router architecture that natively supports parallel two-qubit gate operations, as conceptually illustrated in Fig. \ref{fig1}\textbf{a}. The routing core comprises two physically isolated shared buses (Lane A and Lane B). Interface qubits of distinct modules ($Q_1-Q_4$) are connected to both buses via individual tunable coupling switches ($R_{i\mu}$, where $i\in\{1,2,3,4\}$ and $\mu\in\{A,B\}$). By dynamically programming the switching states, effective pairwise interactions can be mediated on demand through either bus. To formulate the routing protocol, we encode the switch states in a routing configuration matrix that specifies the bus assignment for each interface qubit. For a system with four interface qubits and two shared buses, this configuration is defined as
\[
R=
\begin{pmatrix}
R_{1A} & R_{1B} \\
R_{2A} & R_{2B} \\
R_{3A} & R_{3B} \\
R_{4A} & R_{4B}
\end{pmatrix},
\]
where the matrix elements $R_{i\mu}\in\{0,1\}$ represent the logical states of the individual routing switches:
\[
R_{i\mu}=
\begin{cases}
1, & Q_i \text{ is connected to bus } \mu \text{ via switch } R_{i\mu},\\
0, & Q_i \text{ is isolated from bus } \mu.
\end{cases}
\]
To prevent any individual qubit from participating in dynamical evolutions mediated by both buses simultaneously, we impose the mutual exclusivity constraint:
\[
\sum_{\mu=\mathrm{A},\mathrm{B}} R_{i\mu} \le 1,\quad \forall i\in\{1,\dots,4\}.
\]

Within this framework, parallel CZ gates  correspond to assignments that allocate qubit pairs to distinct buses. For example, to implement a CZ gate between $Q_1$ and $Q_2$, denoted by $\mathrm{CZ}_{12}$, we connect them to Lane A by activating $R_{1\mathrm{A}}$ and $R_{2\mathrm{A}}$, as shown in the second row of the routing table in Table \ref{tb1}. Similarly, to implement $\mathrm{CZ}_{34}$, we selectively connect $Q_3$ and $Q_4$ to Lane B by activating $R_{3\mathrm{B}}$ and $R_{4\mathrm{B}}$, thereby preparing the parallel operation $\mathrm{CZ}_{12} \otimes \mathrm{CZ}_{34}$. For the four-qubit architecture, the complete parallel-gate configuration and routing assignment are shown in Table \ref{tb1}.

Furthermore, based on the routing matrix $R$, we define an effective topological adjacency matrix $A_\mathrm{router}$. This matrix describes whether there exists a bus-mediated physical path between any two qubits: $A_{\mathrm{router}} \equiv RR^\mathrm{T} - \mathrm{diag}\bigl(RR^\mathrm{T}\bigr).$ A nonzero off-diagonal element $\bigl[A_{\mathrm{router}}\bigr]_{ij} = 1$ indicates that qubits $Q_i$ and $Q_j$ are assigned to the same bus.

We note that the routing matrix $R$ is a binary logical description of the routing layer. In an physical circuit-QED implementation, however,the switches are not ideal step-like $0/1$ elements; they are realized by external magnetic fluxes $\Phi_{i\mu}(t)$ that tune the port-coupler frequency and effective coupling strength. Consequently, the matrix elements $R_{i\mu}$ correspond to calibrated ON/OFF control states, rather than directly to a coupling parameter.

\begin{table}[htbp]
\centering
\caption{\textbf{Routing configuration table.} By selectively activating switch combinations, the architecture supports not only isolated two-qubit entangling gates (e.g., $\mathrm{CZ}_{12}$), but also concurrent entangling operations on disjoint qubit pairs (e.g., $\mathrm{CZ}_{12} \otimes \mathrm{CZ}_{34}$) within a single control cycle via bus multiplexing. The routing matrix on the far right rigorously formalizes the underlying hardware connectivity states.}
\footnotesize 
\setlength{\aboverulesep}{0pt}
\setlength{\belowrulesep}{0pt}

\setlength{\tabcolsep}{2pt} 
\setlength{\arraycolsep}{1.5pt}

\renewcommand{\arraystretch}{1.25} 

\begin{tabularx}{\columnwidth}{@{} l >{\centering\arraybackslash}X >{\centering\arraybackslash}X >{\centering\arraybackslash}X >{\centering\arraybackslash}X @{}}
\toprule

\rowcolor{gray!30} 
\rule{0pt}{2.8ex}\textbf{Operation}\rule[-1.2ex]{0pt}{0pt} 
& \textbf{Lane A} 
& \textbf{Lane B} 
& \begin{tabular}[c]{@{}c@{}}\textbf{Active}\\[-1pt]\textbf{switches}\end{tabular} 
& \begin{tabular}[c]{@{}c@{}}\textbf{Routing}\\[-1pt]\textbf{matrix}\end{tabular} \\
\midrule

\rowcolor{white}
Idle & — & — & All OFF & 
$\begin{bmatrix} 0 & 0 & 0 & 0 \\ 0 & 0 & 0 & 0 \end{bmatrix}^{\mathrm{T}}$ \\

\rowcolor{gray!10}
$\mathrm{CZ}_{12}$ & $Q_1, Q_2$ & — & $R_{1\mathrm{A}}, R_{2\mathrm{A}}$ & 
$\begin{bmatrix} 1 & 1 & 0 & 0 \\ 0 & 0 & 0 & 0 \end{bmatrix}^{\mathrm{T}}$ \\

\rowcolor{white}
$\mathrm{CZ}_{34}$ & — & $Q_3, Q_4$ & $R_{3\mathrm{B}}, R_{4\mathrm{B}}$ & 
$\begin{bmatrix} 0 & 0 & 0 & 0 \\ 0 & 0 & 1 & 1 \end{bmatrix}^{\mathrm{T}}$ \\

\rowcolor{gray!10}
$\mathrm{CZ}_{12} \otimes \mathrm{CZ}_{34}$ & $Q_1, Q_2$ & $Q_3, Q_4$ & 
\begin{tabular}[c]{@{}c@{}}$R_{1\mathrm{A}}, R_{2\mathrm{A}}$ \\ $R_{3\mathrm{B}}, R_{4\mathrm{B}}$\end{tabular} & 
$\begin{bmatrix} 1 & 1 & 0 & 0 \\ 0 & 0 & 1 & 1 \end{bmatrix}^{\mathrm{T}}$ \\

\rowcolor{white}
$\mathrm{CZ}_{13} \otimes \mathrm{CZ}_{24}$ & $Q_1, Q_3$ & $Q_2, Q_4$ & 
\begin{tabular}[c]{@{}c@{}}$R_{1\mathrm{A}}, R_{3\mathrm{A}}$ \\ $R_{2\mathrm{B}}, R_{4\mathrm{B}}$\end{tabular} & 
$\begin{bmatrix} 1 & 0 & 1 & 0 \\ 0 & 1 & 0 & 1 \end{bmatrix}^{\mathrm{T}}$ \\

\rowcolor{gray!10}
$\mathrm{CZ}_{14} \otimes \mathrm{CZ}_{23}$ & $Q_1, Q_4$ & $Q_2, Q_3$ & 
\begin{tabular}[c]{@{}c@{}}$R_{1\mathrm{A}}, R_{4\mathrm{A}}$ \\ $R_{2\mathrm{B}}, R_{3\mathrm{B}}$\end{tabular} & 
$\begin{bmatrix} 1 & 0 & 0 & 1 \\ 0 & 1 & 1 & 0 \end{bmatrix}^{\mathrm{T}}$ \\

\bottomrule
\end{tabularx}
\label{tb1}
\end{table}

\subsection{Physical model of Quantum Router}
\label{physical model}
To translate the routing configuration matrix $R$ formulated in Section \ref{Abstract model} into a physical realization, we propose a superconducting quantum router architecture based on circuit quantum electrodynamics (cQED) \cite{PhysRevA.69.062320, cqed, KrantzReview}.

We map the conceptual router model in Fig.~\ref{fig1}\textbf{a} onto the superconducting equivalent circuit shown in Fig.~\ref{fig1}\textbf{b}. In this architecture, the dual shared buses (Lane A and Lane B) are physically implemented by coplanar waveguide (CPW) transmission line resonators, $m_A$ and $m_B$ (shown in blue and red, respectively), which host microwave photonic modes that mediate quantum information exchange. The abstracted port switches $R_{i\mu}$ are physically embodied as SQUID-based nonlinear coupling elements $C_{i\mu}$. Specifically, each interface qubit $Q_i$ is capacitively coupled to two dedicated port couplers $C_{iA}$ and $C_{iB}$, which are in turn capacitively coupled to the bus resonators $m_A$ and $m_B$, respectively. In addition to this coupler-mediated indirect coupling, a direct capacitive coupling pathway is intentionally introduced between the interface qubits and the bus resonators. By dynamically tuning the external magnetic flux bias applied to the couplers, the direct and indirect virtual coupling pathways undergo destructive interference at specific flux operating points, effectively nulling the net coupling in the ``OFF'' state (see Section~\ref{System Hamiltonian}). Consequently, each interface qubit can independently access the shared bus network through two distinct routing channels.

To illustrate how this equivalent circuit is implemented in a physical device, we show in Fig. \ref{fig1}\textbf{c} a conceptual 3D flip-chip design for the modular router. The system is partitioned into a central routing layer and a peripheral qubit layer that comprises four independent computational modules. Each module contains local computational qubits interconnected via nearest-neighbour tunable couplers, which implement intra-module single- and two-qubit operations. To enable quantum information exchange between modules, each module incorporates one or more interface qubits $Q_{i}$. These interface qubits are locally coupled to computational qubits within their respective modules and are connected through a flip-chip architecture to the quantum router located between the modules.

For inter-module two-qubit operations, tuning the relevant port couplers establishes an on-demand, bus-mediated effective interaction between any selected pair of interface qubits. This architecture provides routing redundancy and thereby enhances system-level robustness. It also supports the parallel execution of multiple inter-module two-qubit operations while suppressing residual crosstalk from non-target coupling channels.

\subsection{System Hamiltonian and coupling mechanism}
\label{System Hamiltonian}
We model the router circuit as a network of weakly anharmonic modes comprising four interface qubits, eight port couplers and two bus resonators. Within the Duffing-oscillator and rotating-wave approximations, and setting $\hbar=1$, the full multimode Hamiltonian is
\begin{equation}
\label{system-hami}
\begin{aligned}
\hat{H}
=
\sum_{\lambda\in\mathcal{S}}
\left(
\omega_{\lambda}
\hat{a}_{\lambda}^{\dagger}\hat{a}_{\lambda}
+
\frac{\alpha_{\lambda}}{2}
\hat{a}_{\lambda}^{\dagger}
\hat{a}_{\lambda}^{\dagger}
\hat{a}_{\lambda}
\hat{a}_{\lambda}
\right)
\\ +\sum_{\langle\lambda,\kappa\rangle}
g_{\lambda,\kappa}
\left(
\hat{a}_{\lambda}^{\dagger}\hat{a}_{\kappa}
+
\mathrm{H.c.}
\right).
\end{aligned}
\end{equation}
Here, $\mathcal{S} = \mathcal{Q}\cup\mathcal{C}\cup\mathcal{B}$, where $\mathcal{Q} = \{q_1,q_2,q_3,q_4\}, \mathcal{C} = \left\{c_{i\mu}\,\middle|\,i=1,\ldots,4;\, \mu\in\{\mathrm{A},\mathrm{B}\}\right\}, \mathcal{B} = \{m_{\mathrm{A}},m_{\mathrm{B}}\}.$ The parameters $\omega_{\lambda}$ and $\alpha_{\lambda}$ denote the bare frequency and anharmonicity of mode $\lambda$, respectively. For the linear bus resonators, $\alpha_{m_{\mu}}=0$. The circuit quantization and the relations between the Hamiltonian parameters and the circuit elements are given in Methods.

For each qubit--bus port $(i,\mu)$, the interaction contains a coupler-mediated pathway and a direct capacitive pathway:
\begin{equation}
\begin{aligned}
\hat{H}_{\mathrm{int}}
=
\sum_{i=1}^{4}
\sum_{\mu\in\{\mathrm{A},\mathrm{B}\}}
\Bigl[
&
g_{q_i,c_{i\mu}}
\hat{a}_{q_i}^{\dagger}\hat{a}_{c_{i\mu}}
+
g_{c_{i\mu},m_{\mu}}
\hat{a}_{c_{i\mu}}^{\dagger}\hat{a}_{m_{\mu}}
\\
&+
g_{q_i,m_{\mu}}
\hat{a}_{q_i}^{\dagger}\hat{a}_{m_{\mu}}
+
\mathrm{H.c.}
\Bigr].
\end{aligned}
\end{equation}
The first two terms form the indirect qubit--coupler--bus pathway, whereas the third term describes the direct qubit--bus coupling. The coherent sum of these two pathways provides the interference mechanism used to control each routing channel.

This tunability can be understood perturbatively in the dispersive regime,
\begin{equation}
\label{eqdispersive-condition}
\begin{aligned}
\left|
\Delta_{q_i,c_{i\mu}}
\right|
&\gg
\left|
g_{q_i,c_{i\mu}}
\right|,\\
\left|
\Delta_{m_{\mu},c_{i\mu}}
\right|
&\gg
\left|
g_{c_{i\mu},m_{\mu}}
\right|,
\end{aligned}
\end{equation}
where
\begin{equation}
\begin{aligned}
\Delta_{q_i,c_{i\mu}}
&=
\omega_{q_i}-\omega_{c_{i\mu}},\\
\Delta_{m_{\mu},c_{i\mu}}
&=
\omega_{m_{\mu}}-\omega_{c_{i\mu}}.
\end{aligned}
\label{eq:detunings}
\end{equation}
Eliminating the coupler mode to second order through a Schrieffer--Wolff transformation gives the effective qubit--bus coupling
\begin{equation}
\widetilde{g}_{q_i,m_{\mu}}
=
g_{q_i,m_{\mu}}
+
\frac{
g_{q_i,c_{i\mu}}
g_{c_{i\mu},m_{\mu}}
}{2}
\left(
\frac{1}{\Delta_{q_i,c_{i\mu}}}
+
\frac{1}{\Delta_{m_{\mu},c_{i\mu}}}
\right).
\label{eq:effective-qubit-bus-coupling}
\end{equation}
The first term in Eq.~\eqref{eq:effective-qubit-bus-coupling} is associated with the direct capacitive pathway. The second term describes the coupler-mediated contribution and can be tuned by changing the coupler frequency $\omega_{c_{i\mu}}$. The relative magnitude and sign of the two contributions can therefore be adjusted through the applied flux bias.

We define the idle operating point $\omega_{c_{i\mu}}^{\mathrm{off}}$ through the implicit cancellation condition $\widetilde{g}_{q_i,m_{\mu}}\left(\omega_{c_{i\mu}}^{\mathrm{off}}\right)=0.$ At this operating point, destructive interference suppresses the leading-order transverse exchange interaction between $q_i$ and $m_{\mu}$. Tuning the coupler away from the cancellation point increases
$\lvert\widetilde{g}_{q_i,m_{\mu}}\rvert$ and activates the corresponding routing channel.

Eq.~\eqref{eq:effective-qubit-bus-coupling} is a second-order dispersive approximation that provides a physical interpretation of the switching mechanism. It does not imply that all higher-order or longitudinal interactions vanish exactly at the cancellation point. Accordingly, the idle and active operating frequencies, residual couplings, and gate dynamics reported below are obtained numerically from the full multimode Hamiltonian in Eq.~\eqref{system-hami}, rather than from the perturbative expression alone.

\subsection{ZZ interaction}
\label{cross-kerr interaction}
The realization of CZ gates fundamentally relies on inter-qubit ZZ interactions, which are dynamically engineered via the avoided level crossings between the computational state \(|11\rangle\) and non-computational states such as \(|20\rangle\) or \(|02\rangle\). Based on the 14-mode system Hamiltonian in Eq.~(\ref{system-hami}), we numerically evaluate both the qubit-bus and residual qubit-qubit ZZ couplings across the multi-element network.

To formalize this analysis, the eigenenergies and corresponding eigenstates of the composite circuit are designated as $E_{\mathbf{q},\mathbf{m},\mathbf{c}}$ and
\[
|\mathbf{q},\mathbf{m},\mathbf{c}\rangle
=
|q_1, \dots, q_4, m_A, m_B, c_{1A}, \dots, c_{4B}\rangle,
\]
where the quantum numbers \(q_i\), \(m_\mu\), and \(c_k \in \{0,1,\dots\}\) denote the excitation numbers of the interface qubits, bus resonators, and tunable couplers, respectively. Assuming all couplers remain in their ground states (\(c_k = 0\)), the simplified dressed eigenstates are denoted as \(|\psi_{q_i,m_\mu}\rangle\). The effective ZZ coupling strength between qubit \(Q_i\) and bus resonator \(m_\mu\) is defined as
\[
\zeta_{q_i,m_\mu}
=
E_{1,1}^{q_i,m_\mu}
-
E_{1,0}^{q_i,m_\mu}
-
E_{0,1}^{q_i,m_\mu}
+
E_{0,0}^{q_i,m_\mu},
\]
where \(E_{j,k}^{q_i,m_\mu}\) corresponds to the eigenenergy of the dressed eigenstate \(|\psi_{q_i,m_\mu}\rangle\). To unambiguously establish the correspondence between dressed eigenstates and unperturbed bare product states, we identify each dressed state by maximizing the state overlap:
\[
|\psi_i^{\mathrm{dressed}}\rangle
=
\underset{|\psi_{q_i,m_\mu}\rangle}{\mathrm{argmax}}
\left|
\langle \psi_i^{\mathrm{bare}} \,|\, \psi_{q_i,m_\mu}\rangle
\right|^2.
\]
This state-tracking framework lays the foundation for defining the computational basis at the idling point for subsequent CZ gate synthesis.

\begin{figure}
\centering
\includegraphics[width=1\linewidth]{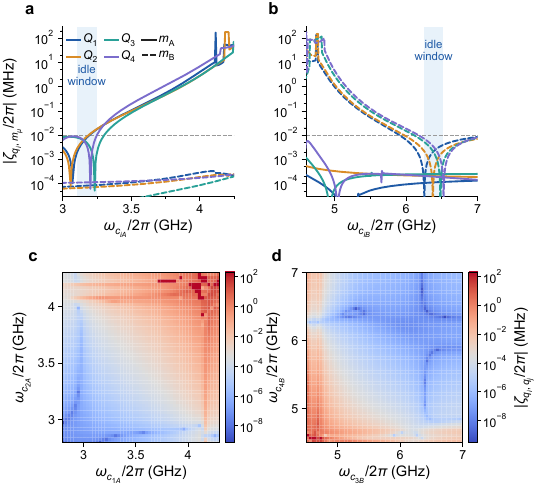}
\caption{
\textbf{Engineering effective ZZ couplings and operating regimes for parallel two-qubit gates.}
\textbf{a,b,} Effective ZZ coupling strengths, $\lvert\zeta_{q_i,m{\mu}}/2\pi\rvert$ (MHz), between the interface qubits $Q_i$ (shown in different colors) and two shared bus resonators, $m_A$ (solid lines) and $m_B$ (dashed lines), as functions of the port-coupler frequency. \textbf{a,} Sweep of the coupler frequency $\omega_{c_{iA}}/2\pi$. \textbf{b,} Sweep of the coupler frequency $\omega_{c_{iB}}/2\pi$. The light-blue shaded regions indicate the predefined idle operating windows, in which $C_{iA}$ operates in the low-frequency range and $C_{iB}$ in the high-frequency range. Within these windows, the effective ZZ couplings between the qubits and the bus resonators are suppressed to the kilohertz level. \textbf{c,d,} Effective interqubit ZZ coupling strengths, $\lvert\zeta_{q_i,q_j}/2\pi\rvert$ (MHz), for parallel two-qubit gate operations, illustrated for the qubit pairs $Q_1-Q_2$ and $Q_3-Q_4$. \textbf{c,} Effective ZZ coupling between $Q_1$ and $Q_2$, mediated by bus $m_A$, as a function of the coupler frequencies $(\omega_{c_{1A}}, \omega_{c_{2A}})$. \textbf{d,} Effective ZZ coupling between $Q_3$ and $Q_4$, mediated by bus $m_B$, as a function of the coupler frequencies $(\omega_{c_{3B}}, \omega_{c_{4B}})$. Dark-blue regions denote decoupled configurations with effective interqubit ZZ couplings below 1~kHz, whereas dark-red regions indicate activated interaction channels with coupling strengths exceeding 1~MHz, suitable for implementing parallel $\mathrm{CZ}_{12} \otimes \mathrm{CZ}_{34}$ gates. All results are obtained by numerically diagonalizing the microscopic full Hamiltonian containing 14 modes and extracting the couplings from the eigenenergies of the dressed states.
}
\label{fig2}
\end{figure}

Fig.~\ref{fig2}\textbf{a} and \ref{fig2}\textbf{b} show the modulation profiles of the effective ZZ coupling rate \(\zeta_{q_i,m_\mu}\) between each qubit \(Q_i\) and the dual shared bus resonators (\(m_A\) and \(m_B\)) as a function of the corresponding port coupler frequency \(\omega_{c_{i\mu}}\). For any given qubit--bus pair, when the coupler is tuned to its idle point, the effective ZZ coupling is substantially suppressed to the \(\mathrm{kHz}\) regime (indicated by the blue shaded regions: solid lines in Fig.~\ref{fig2}\textbf{a} and dashed lines in Fig.~\ref{fig2}\textbf{b}), confirming near-complete decoupling of the interface qubit in its idle state. Conversely, shifting the coupler toward its operating point amplifies the effective ZZ coupling into the \(\mathrm{MHz}\) range, thereby establishing an activated access channel for bus-mediated interactions.

Furthermore, to mitigate inter-bus hybridization crosstalk, we employ an asymmetric detuning scheme where the idle frequencies \(\omega_{c_{iA}}\) of couplers linked to bus \(m_A\) are allocated in the lower frequency band, whereas \(\omega_{c_{iB}}\) associated with bus \(m_B\) are positioned in the higher frequency band. Consequently, when a qubit is routed to a target bus via one port (e.g., \(C_{iA}\) [or \(C_{iB}\)]), the complementary coupler (\(C_{iB}\) [or \(C_{iA}\)]) remains parked at its idling frequency, maintaining the residual ZZ coupling with the non-target bus at a negligible level (as shown by the dashed line in Fig.~\ref{fig2}\textbf{a} and the solid line in Fig.~\ref{fig2}\textbf{b}). These results demonstrate that interface qubits can selectively engage with either bus \(m_A\) or \(m_B\) while concurrently suppressing residual interactions with the idle bus. Therefore, the routing configuration matrix \(R_{i\mu}(t) \in \{0,1\}\) defined in Section~\ref{sec2} is faithfully physically implemented via the programmable switching states of the corresponding couplers.

Building on the demonstrated tunability of individual qubit-bus connections, we examine the inter-qubit ZZ coupling dynamics when two qubits are simultaneously coupled to the same shared bus. As representative examples, we consider the $\mathrm{CZ}_{12}$ gate assigned to Lane A and the $\mathrm{CZ}_{34}$ gate assigned to Lane B, as outlined in Table~\ref{tb1}. Fig.~\ref{fig2}\textbf{c} and \ref{fig2}\textbf{d} show the effective ZZ coupling strengths for the $Q_1-Q_2$ and $Q_3-Q_4$ pairs, respectively, as functions of their corresponding port-coupler frequencies, $\omega_{c_{1A}}$ and $\omega_{c_{2A}}$, and $\omega_{c_{3B}}$ and $\omega_{c_{4B}}$. In the blue regions, the residual inter-qubit ZZ crosstalk is suppressed to the sub-kilohertz level when the couplers are biased at their idle points. In contrast, tuning the couplers into the red regions strongly enhances the ZZ interaction, raising it to the megahertz scale. This clear on/off contrast thereby provides the programmable interaction channels required for subsequent $\mathrm{CZ}$-gate operations.

\begin{table*}[htbp]
\centering
\caption{\textbf{System hardware parameter specifications.} Summary of the Hamiltonian parameters extracted from the circuit setup. The four-qubit system ($Q_1$-$Q_4$) is coupled to two shared bus resonators ($M_A$ and $M_B$, corresponding to Lane A and Lane B, respectively) via frequency-tunable couplers ($C_{i\mathrm{A}}$ and $C_{i\mathrm{B}}$).}
\footnotesize 
\setlength{\aboverulesep}{0pt}
\setlength{\belowrulesep}{0pt}

\setlength{\tabcolsep}{6pt} 
\setlength{\arraycolsep}{2pt}

\renewcommand{\arraystretch}{1.25} 

\begin{tabularx}{\textwidth}{@{} l l c c >{\centering\arraybackslash}X >{\centering\arraybackslash}X >{\centering\arraybackslash}X >{\centering\arraybackslash}X @{}}
\toprule

\rowcolor{gray!30} 
\rule{0pt}{2.8ex}\textbf{Subsystem} 
& \textbf{Parameter} 
& \textbf{Symbol} 
& \textbf{Unit} 
& \textbf{$Q_1$} 
& \textbf{$Q_2$} 
& \textbf{$Q_3$} 
& \textbf{$Q_4$} \rule[-1.2ex]{0pt}{0pt} \\
\midrule

\rowcolor{white}
\textbf{Qubits} & Operating frequency & $\omega_{q_i} / 2\pi$ & GHz & 4.100 & 4.200 & 4.450 & 4.560 \\

\rowcolor{gray!10}
& Anharmonicity & $\alpha_{q_i} / 2\pi$ & MHz & $-324$ & $-280$ & $-300$ & $-314$ \\

\rowcolor{white}
\textbf{Couplers} & Idle frequency (Lane A) & $\omega_{c_{iA}} / 2\pi$ & GHz & 3.100 & 3.090 & 3.140 & 3.100 \\

\rowcolor{gray!10}
& Idle frequency (Lane B) & $\omega_{c_{iB}} / 2\pi$ & GHz & 6.375 & 6.420 & 6.525 & 6.565 \\

\rowcolor{white}
& Anharmonicity & $\alpha_{c_{i\mu}} / 2\pi$ & MHz & \multicolumn{4}{c}{$-300.0$ (uniform for all $C_{i\mathrm{A}}$ and $C_{i\mathrm{B}}$)} \\

\rowcolor{gray!10}
\textbf{Bus Resonators} & Bare resonance frequency & $\omega_{m_{\mu}} / 2\pi$ & GHz & \multicolumn{2}{c}{Bus 1 ($m_A$, Lane A): 4.300} & \multicolumn{2}{c}{Bus 2 ($m_B$, Lane B): 4.750} \\

\rowcolor{white}
\textbf{Couplings} & Qubit--coupler coupling & $g_{q_i,c_{i\mu}} / 2\pi$ & MHz & 84.0 & 84.0 & 84.0 & 84.0 \\

\rowcolor{gray!10}
& Coupler--bus coupling & $g_{c_{i\mu},m_{\mu}} / 2\pi$ & MHz & \multicolumn{2}{c}{$g_{c_{iA},m_A} = -62.0$ (Lane A)} & \multicolumn{2}{c}{$g_{c_{iB},m_B} = +84.0$ (Lane B)} \\

\rowcolor{white}
& Coupling to Bus 1 & $g_{q_i,m_A} / 2\pi$ & MHz & 4.1 & 3.9 & 4.3 & 4.3 \\

\rowcolor{gray!10}
& Coupling to Bus 2 & $g_{q_i,m_B} / 2\pi$ & MHz & 4.3 & 4.1 & 4.1 & 4.1 \\

\bottomrule
\end{tabularx}
\label{tb:hardware_params}
\end{table*}

\subsection{CZ Gate performance}
\label{CZ gate}
The two-qubit \(\mathrm{CZ}\) gates are realized by synchronously tuning the couplers that connect the target qubits to the same bus resonator. In the following simulations, we use the full system Hamiltonian, rather than an effective-Hamiltonian model. All parameters are listed in Table~\ref{tb:hardware_params}, and the time evolution is obtained using the QuTiP solver \cite{JOHANSSON20121760}.

We first define the computational basis states \(\ket{i,j,k,l}\), with \(i,j,k,l\in\{0,1\}\), as the eigenstates of the full circuit Hamiltonian when all tunable couplers are biased at their idle points. For notational simplicity, modes outside the computational subspace are omitted. 

\begin{figure*}
\centering
\includegraphics[width=1\linewidth]{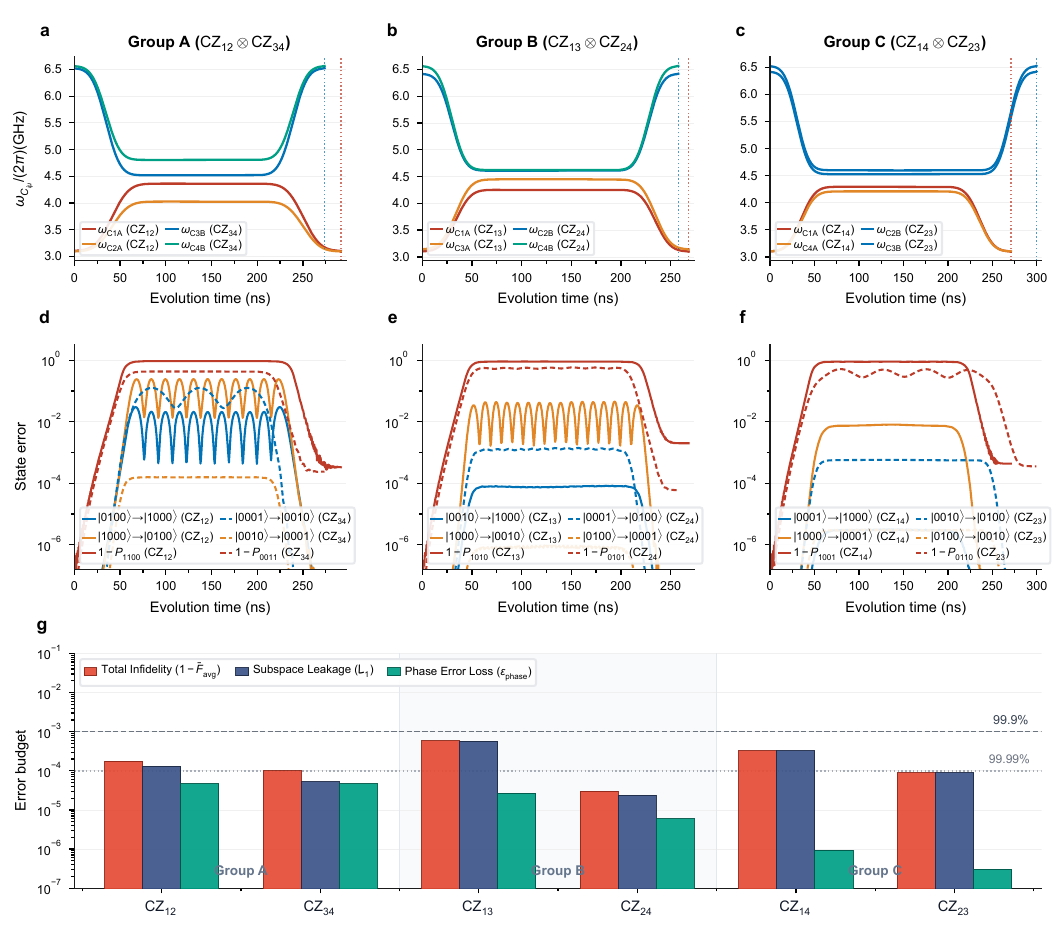}
\caption{\textbf{CZ-gate pulses and system dynamics.}
\textbf{a--c,} Coupler-frequency modulation trajectories, $\omega_C(t)$, for three groups of parallel two-qubit CZ gates: Group A, $\mathrm{CZ}_{12}\otimes\mathrm{CZ}_{34}$; Group B, $\mathrm{CZ}_{13}\otimes\mathrm{CZ}_{24}$; and Group C, $\mathrm{CZ}_{14}\otimes\mathrm{CZ}_{23}$. The pulse waveforms are self-consistently compensated for differences in gate duration; for example, $T_{12}<T_{34}$ in \textbf{a}. Dashed segments indicate the waiting intervals during which a coupler that has completed its gate operation first is returned to and held at its idle frequency until the end of the parallel operation cycle. This synchronized scheduling preserves a common global clock and avoids the timing gaps that would arise from asynchronous execution of pulses with unequal durations.
\textbf{d--f,} Log-scale state-error dynamics obtained by isolating each CZ gate while using the synchronized pulse parameters described above, with all non-target qubits held at their idle frequencies. The curves resolve transient population transfer between computational and noncomputational states, transient population exchange between computational states, such as $\lvert 0100\rangle \leftrightarrow \lvert 1000\rangle$, and computational-state population errors quantified by $1-P_{\psi}$.
\textbf{g,} Quantitative error decomposition for the isolated CZ gates. The bars show the average gate infidelity $1-F_{\mathrm{avg}}$ (red), computational-subspace leakage $L_1$ (dark blue), and conditional-phase error $\epsilon_{\mathrm{phase}}$ (teal). The two horizontal dashed lines mark $1-F_{\mathrm{avg}}=10^{-3}$ ($F_{\mathrm{avg}}=99.9\%$) and $1-F_{\mathrm{avg}}=10^{-4}$ ($F_{\mathrm{avg}}=99.99\%$), respectively, as a commonly used gate-fidelity benchmark and a more stringent high-fidelity target.
}
\label{fig3}
\end{figure*}

For a parallel two-qubit operation composed of two independent gates, such as \(\mathrm{CZ}_{12}\otimes \mathrm{CZ}_{34}\), the optimized parameter sets of the individual branches form a high-dimensional initial parameter space for joint optimization of the parallel gate. In general, the optimal gate durations for the two individual CZ branches are asymmetric. Thus, the total duration of the parallel gate is determined by  $T_{\mathrm{parallel}}=\max(T_{12},T_{34})$. For the branch with the shorter gate duration, once the target dynamical evolution has been completed, the corresponding coupler pulse is smoothly returned to its idle point and kept there until the end of the synchronized gate sequence. As shown in Fig.~\ref{fig3}\textbf{a-c}, the pulse sequences during the synchronized \(\mathrm{CZ}\)-gate operation include dashed segments that denote such waiting intervals. This asynchronous padding procedure reflects the practical timing logic of experimental control sequences.

Building on this control protocol, we first performed an exact simulation of the dynamics induced by the modulation pulses under isolated-operation conditions, with all non-target qubits initialized in the ground state \(|0\rangle\). Fig.~\ref{fig3}\textbf{d-f} show the corresponding error dynamics for three representative configurations. During the pulse evolution, the computational subspace exhibits population oscillations, including SWAP-type errors such as \(|0100\rangle \leftrightarrow |1000\rangle\) and \(|0001\rangle \leftrightarrow |0010\rangle\) in Group~A, together with transient leakage from the target states, quantified by \(1-P_{|1100\rangle}\) and \(1-P_{|0011\rangle}\). These excitations, however, are recovered back into the computational subspace during the pulse-turnoff stage. Fig.~\ref{fig3}\textbf{g} shows that the average gate infidelity \(1-F_{\mathrm{avg}}\) of the isolated CZ gates remains below \(10^{-3}\) for all cases, corresponding to \(F_{\mathrm{avg}}>99.9\%\), and reaches \(F_{\mathrm{avg}}>99.99\%\) for some qubit pairs, such as \(\mathrm{CZ}_{24}\) and \(\mathrm{CZ}_{23}\). Further decomposing the error into the computational-subspace leakage rate \(L_1\) and the conditional phase error \(\varepsilon_{\mathrm{phase}}\), we find that the infidelity is dominated by transient leakage out of the subspace, while both metrics remain strongly suppressed. These results confirm the capability of the platform to achieve high-fidelity isolated-gate control.

\begin{figure}
\centering
\includegraphics[width=1\linewidth]{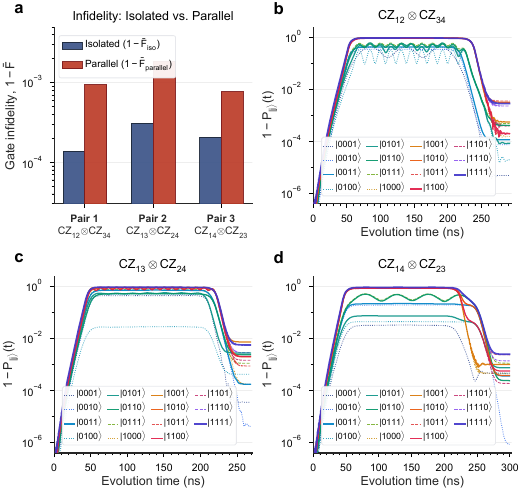}
\caption{
\textbf{Fidelities and state dynamics of parallel CZ gates.} 
\textbf{a,} Quantitative comparison between the isolated average infidelities (blue bars, $1 - \bar{F}_{\mathrm{iso}}$) and the parallel gate infidelities (red bars, $1 - F_{\mathrm{parallel}}$) for three representative pairing configurations (Pair 1: $\mathrm{CZ}_{12} \otimes \mathrm{CZ}_{34}$; Pair 2: $\mathrm{CZ}_{13} \otimes \mathrm{CZ}_{24}$; Pair 3: $\mathrm{CZ}_{14} \otimes \mathrm{CZ}_{23}$). 
\textbf{b--d,} Dynamics of the state errors, $1 - P_{|j\rangle}(t)$, for all 15 non-ground computational basis states, $|j\rangle \in \{|0001\rangle, |0010\rangle, \dots, |1111\rangle\}$, driven by the synchronized pulse sequences for the three configurations. During the pulse flat-top period, many-body interactions induce transient coherent population transfers and excitations within the computational subspace. Following the smooth ramp-down of the pulses, the errors for the vast majority of the basis states rapidly fall back to the $10^{-3}$--$10^{-5}$ level. Compared with the other two configurations, the parallel gate error for Pair 2 ($\mathrm{CZ}_{13} \otimes \mathrm{CZ}_{24}$) exhibits a slight increase ($1.8 \times 10^{-3}$). As revealed by the resolved trajectories at the end of the time evolution in \textbf{c}, this excess error originates primarily from weak non-adiabatic transitions involving the multi-excitation states $|1001\rangle$ and $|1111\rangle$ under the synchronized modulation.
}
\label{fig4}
\end{figure}

We then activated all drive channels simultaneously in the full Hamiltonian dynamics model to systematically evaluate the fidelity of parallel gate execution. Fig.~\ref{fig4}\textbf{a} compares, for three parallel connectivity patterns -- Pair~1: $\mathrm{CZ}_{12}\otimes \mathrm{CZ}_{34}$, Pair~2: $\mathrm{CZ}_{13}\otimes \mathrm{CZ}_{24}$, and Pair~3: $\mathrm{CZ}_{14}\otimes \mathrm{CZ}_{23}$ -- the average added isolated-gate infidelity, $1-\bar{F}_{\mathrm{iso}}= ({(1-\bar{F}_{ij})+(1-\bar{F}_{kl})})/2$, with the average parallel-gate infidelity, $1-F_{\mathrm{parallel}}$. The numerical results show that the three parallel CZ operations exhibit infidelities of $9.4\times10^{-4}$, $1.8\times10^{-3}$, and $7.9\times10^{-4}$, respectively. The additional distortion induced by parallel execution, $\Delta(1-\bar{F})$, remains strictly bounded at the level $10^{-3}$.

To characterize the coherent dynamics during parallel entangling-gate execution, Fig.~\ref{fig4}\textbf{b-d} track the transient population errors, $1-P_{|j\rangle}(t)$, for the 15 non-vacuum computational basis states $|j\rangle\in\{|0001\rangle, |0010\rangle, \ldots, |1111\rangle\}$ over the gate duration. During the flat-top stage of the modulation pulse, the target-state populations are broadly transferred to non-computational intermediate manifolds to accumulate the desired phase. During the subsequent adiabatic turn-off stage, all participating entangled-state populations coherently refocus along closed trajectories back to their respective computational basis states. By the end of the gate duration, the residual population errors of most computational states have decayed to the $10^{-3}$--$10^{-5}$ range. Compared with the other two configurations, Pair~2 ($\mathrm{CZ}_{13}\otimes \mathrm{CZ}_{24}$) exhibits a slightly larger parallel-gate infidelity of $1.8\times10^{-3}$. Trajectory decomposition at the end of the dynamics in Fig.~\ref{fig4}\textbf{c} indicates that this excess error arises primarily from weak nonadiabatic transitions involving the multi-excitation states $|1001\rangle$ and $|1111\rangle$ under synchronized modulation.

\subsection{Decoherence-Limited Fidelity of Parallel CZ Gates}
\label{decoherence}
Although closed-system simulations accurately capture coherent control errors and state leakage, the ultimate fidelity of the parallel $\mathrm{CZ}$ gates is fundamentally constrained by energy and dephasing relaxation of the physical qubits. In this section, we quantify the impact of finite coherence times on the performance of parallel gate operations.

\begin{figure*}
\centering
\includegraphics[width=1\linewidth]{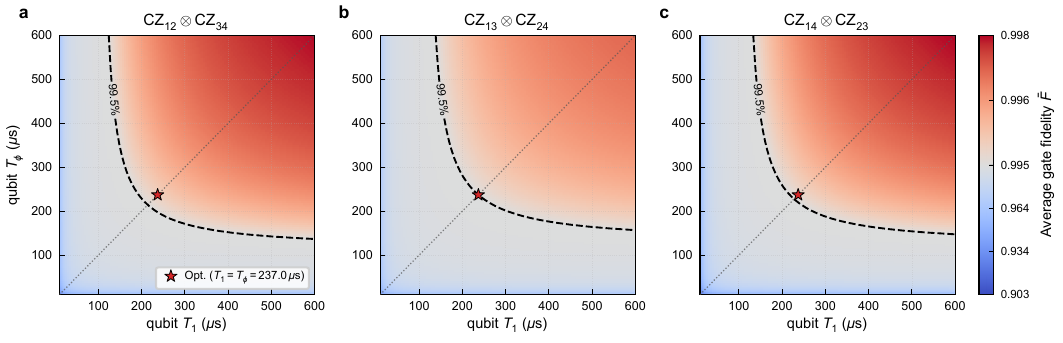}
\caption{\textbf{Impact of qubit decoherence times on parallel two-qubit gate fidelities.} 
\textbf{a}--\textbf{c}, 2D parameter landscapes of the simulated average gate fidelity $\bar{F}$ as a function of energy relaxation time $T_1$ and pure dephasing time $T_\phi$ for three parallel configurations: $\mathrm{CZ}_{12}\otimes\mathrm{CZ}_{34}$ \textbf{a}, $\mathrm{CZ}_{13}\otimes\mathrm{CZ}_{24}$ \textbf{b}, and $\mathrm{CZ}_{14}\otimes\mathrm{CZ}_{23}$ \textbf{c}. In each panel, uniform coherence parameters are assumed across all participating interface qubits throughout the parameter sweep. The dashed black contours delineate the $\bar{F} = 99.5\%$ fidelity threshold boundary, while the grey dotted diagonal lines represent the symmetric coherence condition ($T_1 = T_\phi$). The red star marks the critical coherence benchmark, defining the minimum symmetric coherence time required to simultaneously surpass the $99.5\%$ fidelity threshold across all parallel gate configurations.
}
\label{fig5}
\end{figure*}

In the framework of open quantum systems, decoherence is rigorously formulated via Lindblad dissipators. For a specific jump operator $\hat{L}_k$, the corresponding dissipator acting on the density matrix of the system $\hat{\rho}$ is defined as
\begin{equation}
\mathcal{D}[\hat{L}_k]\hat{\rho} = \hat{L}_k \hat{\rho} \hat{L}_k^{\dagger} - \frac{1}{2} \left( \hat{L}_k^{\dagger}\hat{L}_k \hat{\rho} + \hat{\rho}\hat{L}_k^{\dagger}\hat{L}_k \right).
\end{equation}
For $N_L$ independent dissipation channels, the overall time evolution of the system is governed by the Lindblad master equation:
\begin{equation}
\dot{\hat{\rho}}(t) = -\frac{i}{\hbar} \left[ \hat{H}(t),\hat{\rho}(t) \right] + \sum_{k=1}^{N_L} \Gamma_k \mathcal{D}[\hat{L}_k]\hat{\rho}(t),
\end{equation}
where $\Gamma_k$ dictates the decoherence rate associated with the $k$-th dissipation process \cite{PhysRevLett.129.150504,Abad2025impactofdecoherence}.

In our architecture, we primarily account for the energy relaxation and pure dephasing of the four physical qubits. Specifically, for the $j$-th qubit, energy relaxation is modeled by the jump operator $\hat{L}_{1,j} = \hat{a}_j$ with a corresponding decay rate $\Gamma_{1,j} = 1/T_{1,j}$, where $T_{1,j}$ is the relaxation time. Concurrently, pure dephasing is described by the jump operator $\hat{L}_{\phi,j} = \sqrt{2}\,\hat{a}_j^{\dagger}\hat{a}_j$ with the dephasing rate $\Gamma_{\phi,j} = 1/T_{\phi,j}$.

By expanding the master equation to first order in the dissipation rates, the leading-order impact of the incoherent channels on the average gate fidelity is expressed as:
\begin{equation}
\bar{F} \approx 1 + \sum_{k=1}^{N_L} \Gamma_k \int_{0}^{\tau} dt\,\delta F_k(t) + \mathcal{O}(\tau^2\Gamma_k^2),
\end{equation}
where $\tau$ is the gate duration, and $\delta F_k(t)$ quantifies the instantaneous fidelity penalty induced by the $k$-th dissipation channel. For a given jump operator $\hat{L}_k$, this instantaneous contribution takes the form:
\begin{equation}
\begin{aligned}
\delta F_k(t) = \frac{1}{d(d+1)} \left| \mathrm{Tr}_{\mathrm{cmp}} \left[ \hat{L}_k(t) \right] \right|^2 \\ - \frac{1}{d+1} \mathrm{Tr}_{\mathrm{cmp}} \left[ \hat{L}_k^{\dagger}(t)\hat{L}_k(t) \right],
\end{aligned}
\end{equation}
where $\hat{L}_k(t) \equiv \hat{U}^{\dagger}(t)\hat{L}_k\hat{U}(t)$ represents the jump operator in the interaction picture. For the parallel two-qubit gates evaluated here, the computational subspace spans a dimension of $d=16$, and $\mathrm{Tr}_{\mathrm{cmp}}[\cdot]$ denotes the trace restricted to this computational manifold.

Based on the aforementioned first-order perturbative master equation framework, we evaluate the average gate fidelity $\bar{F}$ across a two-dimensional coherence parameter space ($T_1, T_\phi \in [10, 600]\,\mu\mathrm{s}$) for the three parallel gate configurations: \textbf{a} $\mathrm{CZ}_{12}\otimes\mathrm{CZ}_{34}$, \textbf{b} $\mathrm{CZ}_{13}\otimes\mathrm{CZ}_{24}$, and \textbf{c} $\mathrm{CZ}_{14}\otimes\mathrm{CZ}_{23}$, as depicted in Fig.~\ref{fig5}. To facilitate direct comparison, all panels share a unified colormap and delineate the chosen fidelity target of $\bar{F} = 99.5\%$ (black dashed contours). The horizontal and vertical axes correspond to the qubit energy relaxation time $T_1$ and pure dephasing time $T_\phi$, respectively, assuming homogeneous decoherence rates across the four qubits.

To determine the minimum coherence budget required to simultaneously drive all three parallel configurations above the representative high-fidelity benchmark of ($\bar{F} \ge 99.5\%$) under contemporary superconducting hardware conditions, we formulate a constrained optimization problem over the joint fidelity landscape:
\begin{equation}
\begin{aligned}
\min_{(T_1, T_\phi)} \quad & \sqrt{T_1^2 + T_\phi^2} \\
\text{s.t.} \quad & \min_{k\in\{a,b,c\}} \bar{F}_k(T_1, T_\phi) \ge 99.5\%.
\end{aligned}
\end{equation}
Under the balanced noise assumption ($T_1 = T_\phi$) commonly adopted in superconducting processor benchmarks, the critical coherence time required to universally attain $\bar{F} \ge 99.5\%$ across all synchronized operations is $T_1 = T_\phi = 237.00\,\mu\mathrm{s}$ (marked by red stars in each panel). This result establishes a concrete hardware coherence target for parallel gate execution, which is well within experimental reach for state-of-the-art fixed-frequency transmon architectures.

\subsection{Algorithm-level benefits of reconfigurable routing}
\label{Algorithm}
\begin{figure}
\centering
\includegraphics[width=1\linewidth]{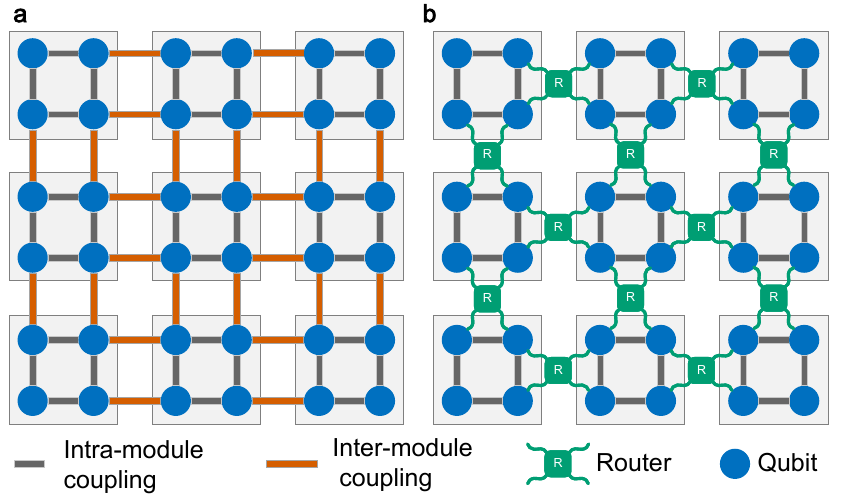}
\caption{
\textbf{Scalable topological architectures for modular quantum processors.}
\textbf{a,} Chip topology scaled via a conventional grid structure. The processor comprises an array of computational modules (light-grey squares). Within each module, four qubits (blue circles) are interconnected by intra-module couplers (grey lines) to form a local sub-grid. This four-qubit configuration represents the minimal non-trivial unit capable of simultaneously accommodating local connectivity, inter-module routing, and parallel two-qubit gate operations. Adjacent modules are directly linked via one-to-one inter-module couplers (orange lines) connecting their peripheral qubits, thereby forming a global two-dimensional grid lattice. 
\textbf{b,} Chip topology scaled via a router-based architecture. In contrast to the direct hard-wiring scheme in \textbf{a}, inter-module connectivity between adjacent computational modules is mediated by quantum routers (denoted by green `R' nodes). By replacing point-to-point direct couplings with multi-port router nodes, this architecture provides significant topological flexibility for inter-module entanglement and information routing while preserving global connectivity.
}
\label{fig6}
\end{figure}

\begin{figure*}
\centering
\includegraphics[width=0.98\linewidth]{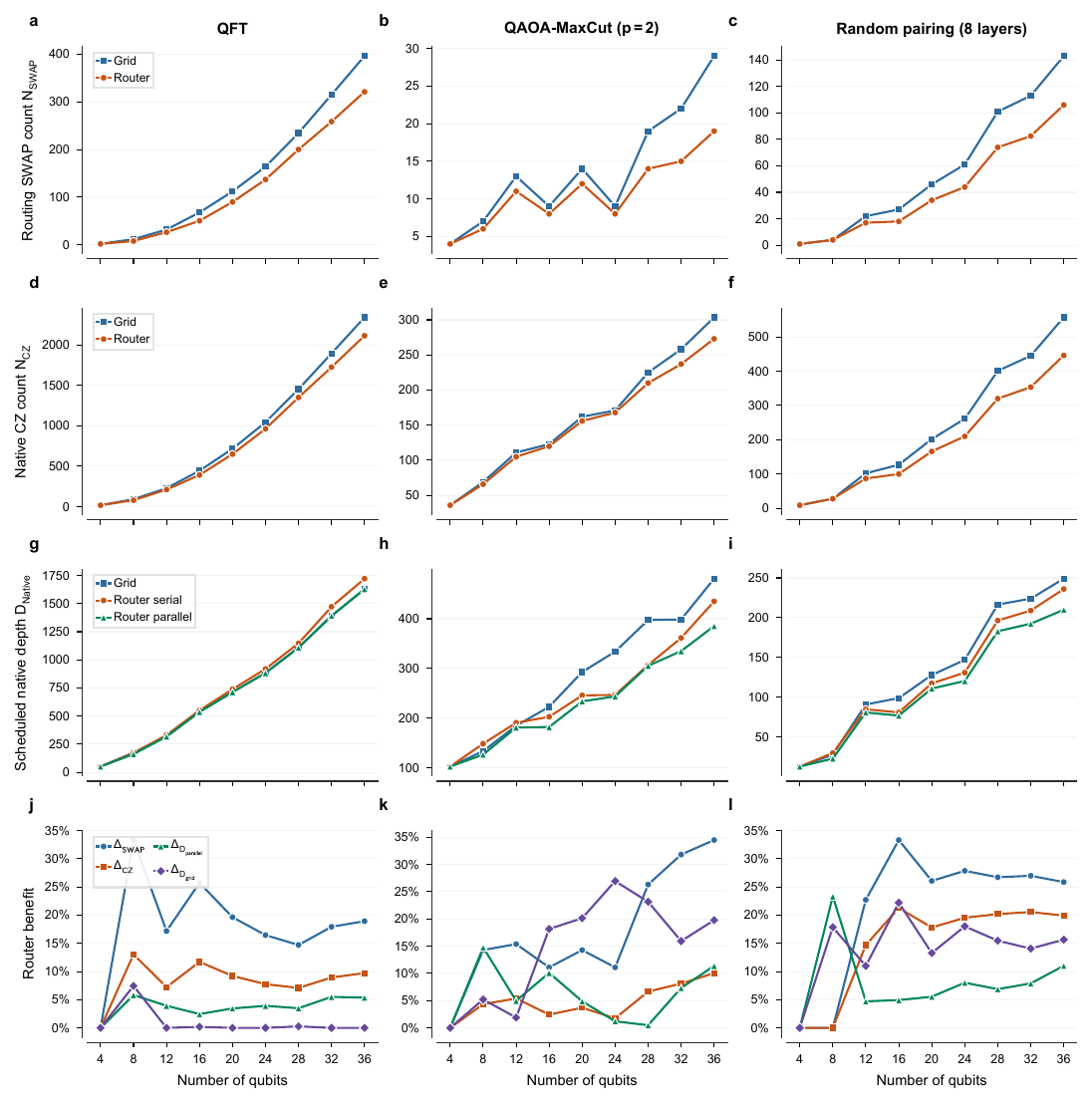}
\caption{\textbf{Compilation overhead and performance gains of a modular quantum-router architecture for representative quantum algorithms.}
\textbf{a--c,} Scaling of the total number of additional SWAP gates introduced during physical routing as a function of the number of qubits ($n=4\text{--}36$) for three representative benchmark circuits: the quantum Fourier transform (\textbf{a,} QFT), the quantum approximate optimization algorithm for MaxCut (\textbf{b}, QAOA-MaxCut, $p=2$), and an eight-layer random-pairing circuit (\textbf{c}, Random pairing (8 layers)). The results compare the hardware-mapping overhead of a conventional two-dimensional nearest-neighbour grid topology (Grid, blue squares) with that of the modular quantum-router architecture proposed here (Router, orange circles).
\textbf{d--f,} Scaling of the total gate count after decomposing the circuits into hardware-native controlled-$Z$ gates (native CZ count). The reduction in routing-induced SWAP gates enabled by the router topology translates directly into a substantial reduction in the number of native two-qubit gates.
\textbf{g--i,} Scheduled native-gate depth obtained using an as-soon-as-possible (ASAP) scheduling algorithm. The results compare the grid topology (Grid, blue squares), the serial router mode (Router serial, orange circles), and the parallel router mode (Router parallel, green triangles). In parallel mode, non-conflicting inter-module CZ gates acting on distinct qubits and not sharing router-control resources can be executed in the same scheduling layer.
\textbf{j--l,} Relative performance gains provided by the quantum-router architecture over the conventional grid topology. The blue circles denote the SWAP-gate reduction $\Delta_{\mathrm{SWAP}}$, the orange squares denote the native-CZ-gate reduction $\Delta_{\mathrm{CZ}}$, the green triangles denote the reduction in circuit depth achieved by parallel scheduling relative to serial scheduling, $\Delta_{D_{\mathrm{parallel}}}$, and the purple diamonds denote the total circuit-depth reduction of the parallel router mode relative to the grid topology, $\Delta_{D_{\mathrm{grid}}}$. All circuits were mapped using the same gate-decomposition rules and the SABRE heuristic compilation strategy. Each data point represents the median compilation result over multiple random seeds.}
\label{fig7}
\end{figure*}

As established previously at physical level, the dual-bus router architecture inherently supports parallel CZ gate execution. Here, we investigate whether this physical hardware capability translates into tangible algorithmic benefits. To this end, we compile three representative benchmark circuits---QFT, QAOA-MaxCut, and random quantum circuits---onto both the Grid and Router topologies (as depicted in Fig.~\ref{fig6}). We then evaluate and compare the resultant SWAP count, native CZ gate count, and overall circuit depth. The performance curves in Fig.~\ref{fig7} represent the statistical medians derived from 100 random SABRE seeds.

Both topological architectures utilize a 4-qubit grid as their fundamental building block. This specific choice does not imply that a four-qubit block is the unique or optimal module scale for modular quantum processors. Rather, it is designed to serve as a minimal yet sufficiently complex baseline unit---one that accommodates local connectivity, multiple inter-module interfaces, and the capacity for two disjoint two-qubit operations simultaneously. During system scaling, the intra-module connectivity and interface definitions remain strictly invariant; only the total number of modules is increased. Consequently, any observed performance disparities between the two architectures can be primarily attributed to the inter-module routing strategies, rather than variations in the internal module structure.

In our compilation model, the Router topology retains all inter-module connections natively present in the Grid topology, while additionally incorporating the cross-module connections facilitated by the router hardware. Mathematically, this dictates that the Grid topology is a subgraph of the Router topology. This structural design guarantees that both architectures possess identical qubit counts and intra-module layouts, ensuring that the ensuing comparative analysis directly and unambiguously isolates the impacts of the enhanced connectivity and parallel routing resources.

As illustrated in Fig.~\ref{fig7}\textbf{a-c}, owing to the enhanced cross-module connectivity, the Router topology significantly reduces the number of inserted SWAP gates across all tested circuits, a topological advantage that scales favorably with system size. Furthermore, because each logical SWAP operation must be physically decomposed into multiple native two-qubit gates, this algorithmic-level routing optimization translates directly into a substantial reduction in the overall native CZ gate count [Fig.~\ref{fig7}\textbf{d-f}]. For instance, in a larger-scale 36-qubit system, the Router topology mitigates the SWAP overhead by approximately 19\%--34\% and yields a 10\%--20\% reduction in native CZ execution overhead across the three benchmark circuit types.

Importantly, a reduction in gate count does not necessarily guarantee a proportional decrease in circuit depth. Circuit depth is strictly governed by the critical dependency path: if the eliminated gates were originally scheduled in parallel with other operations, their removal may have a negligible impact on the overall depth. Significant depth reduction occurs only when operations or idle waiting layers along the critical path are truncated. This distinction is particularly evident in the QFT benchmark. For the 36-qubit QFT, parallel scheduling reduces the depth by approximately 4.9\% compared to serial routing, yet it yields a net-zero end-to-end depth improvement over the Grid baseline, as shown in Fig.~\ref{fig7}\textbf{g}. In essence, while parallel execution successfully compensates for the waiting times induced by serial routing constraints, it fails to further compress the fundamental critical path inherent to the Grid. This arises because a large portion of the controlled-phase gates in QFT are highly constrained by shared qubits and strict sequential dependencies, inherently limiting the number of cross-module gates that can be executed concurrently.

Conversely, QAOA-MaxCut and random pairing circuits feature a higher density of independent two-qubit operations, allowing them to more effectively exploit the augmented connectivity and parallel resources. At the 36-qubit scale, as shown in Fig.~\ref{fig7}\textbf{h,i}, the median end-to-end depth reductions for QAOA-MaxCut and random circuits relative to the Grid baseline are approximately 20\% and 15\%, respectively. Notably, the random circuit under parallel scheduling achieves an additional 10\% depth reduction over its serial counterpart, exhibiting the most pronounced advantage of parallel scheduling among the three benchmarks.

As observed in Fig.~\ref{fig7}\textbf{j-l}, the relative performance gains do not exhibit strictly monotonic scaling behavior with respect to system size. SABRE is fundamentally a heuristic mapping algorithm. Different system scales correspond to distinct logical interaction graphs, initial qubit layouts, and discrete gate dependencies. Consequently, at certain scales, the heuristic may incidentally converge on layouts where the Grid architecture performs comparably to, or marginally better than, the Router architecture. Furthermore, the extent to which parallel scheduling can compress circuit depth depends strictly on the number of independent CZ gates per layer that simultaneously satisfy logical dependencies, qubit disjointness, and router capacity constraints. Therefore, the localized fluctuations observed in the plots are inherent characteristics of discrete compilation and scheduling problems.

Overall, this comparative analysis elucidates two complementary roles of the Router architecture. The addition of cross-module connections primarily serves to reduce the number of SWAP and native CZ gates, whereas parallel routing resources are primarily responsible for minimizing idle waiting times between independent cross-module gates. The former benefit is largely dictated by the logical interaction distances within the algorithm, while the latter heavily relies on the intrinsic parallel structure of the circuit itself. Consequently, while the hardware connectivity and parallel execution capabilities of the router successfully translate into algorithmic-level improvements, the ultimate gains in circuit depth exhibit explicit, circuit-specific dependencies.

\section{Discussion}\label{sec3}

In this work, we propose and systematically evaluate a bus-based reconfigurable quantum router architecture for modular superconducting quantum processors, establishing a cross-stack analytical framework that spans from system Hamiltonians and pulse-level coherent control to algorithmic compilation and scheduling. The core principle of this router lies in exploiting the destructive interference mechanism between multiple shared buses and a tunable SQUID network. This allows for the on-demand establishment of effective cross-module couplings while strictly suppressing residual interactions in non-target channels. Consequently, the quantum router evolves beyond a mere intermediary connecting element into a schedulable hardware resource capable of supporting parallel two-qubit operations.

At the physical level, appropriate flux biasing induces destructive interference between direct capacitive couplings and coupler-mediated indirect pathways, suppressing both effective couplings and residual ZZ interactions in idle channels to the kilohertz regime. The adoption of distinct idle frequency windows for the two buses further mitigates unintended cross-channel couplings. Full Hamiltonian dynamical simulations demonstrate that two disjoint CZ gates can be executed simultaneously, maintaining gate errors on the order of $10^{-3}$. Error analysis reveals that the remaining infidelity is predominantly limited by leakage out of the computational subspace, pointing to the primary direction for further gate optimization. Furthermore, open-system analysis delineates the design requirements for device coherence. Assuming uniform relaxation and dephasing times across the four interface qubits, achieving an average fidelity of $\bar{F} \ge 99.5\%$ simultaneously across three synchronous CZ configurations demands a symmetric coherence time of approximately $T_1 = T_\phi = 237\ \mu\text{s}$. This value should be interpreted as a device design benchmark under the current model, rather than a universal experimental threshold, as real-world implementations will be further subject to bus losses, coupler decoherence, thermal photons, and control noise.

At the algorithmic and compilation level, the router yields two complementary benefits. First, the enhanced cross-module connectivity shortens the routing paths for specific logical qubits, thereby reducing the number of SWAP gates---and their decomposed native CZ gates---in QFT, QAOA-MaxCut, and random pairing circuits. Second, the dual-channel resources permit the synchronous execution of cross-module CZ gates that satisfy logical dependencies and qubit disjointness, thereby minimizing wait times along the critical path. Notably, these two benefits are not equivalent: the reduction in gate count is primarily driven by improved connectivity, whereas depth compression relies heavily on the intrinsic parallel structure of the circuit itself. Random pairing circuits, which are rich in independent two-qubit operations, can fully exploit the synchronous routing resources. QAOA-MaxCut also demonstrates end-to-end depth reductions but exhibits higher sensitivity to the specific SABRE mapping. In contrast, while QFT benefits from reduced SWAP and CZ gate counts, its strict logical dependencies severely restrict critical path compression. This underscores the fact that auxiliary connectivity does not automatically translate into proportional depth reductions. The system-level superiority of the router demands a synergy among hardware connectivity, gate fidelity, and compiler scheduling.

It is imperative to note that while theoretical and numerical validations demonstrate the superiority of this architecture, translating it into a practical device faces several engineering challenges. First, while providing long-range connectivity, the central buses may introduce Purcell decay, residual thermal photons, and frequency crowding. These issues must be rigorously managed through dedicated filtering, careful frequency allocation, and advanced layout engineering. Second, as the module scale expands, routing capabilities will eventually be constrained by the finite number of bus lanes and parallel scheduling conflicts. Therefore, the dual-bus topology discussed herein is best viewed as a minimal parallel-routable unit. For larger-scale processors, this architecture can be logically extended into multi-lane routing layers or integrated into hierarchical modular networks.

Overall, this work demonstrates that the true value of a reconfigurable quantum router extends beyond merely ``connecting more qubits''; it lies in organizing these connections into parallel resources that can be explicitly scheduled by compilers. The ultimate algorithmic payoff is jointly determined by the low residual coupling and synchronous gate capabilities at the physical layer, combined with resource-aware scheduling at the compilation layer. This framework provides a verifiable design roadmap for constructing modular superconducting quantum processors characterized by high connectivity, low crosstalk, and exploitable parallelism.

\section{Methods}
\label{sec4}
\subsection{Circuit Quantization}\label{quantization}
\subsubsection{System Lagrangian}
\label{Lagrangian}
Fig.~\ref{fig1} shows the circuit model in which the qubits are coupled to the router. The self-capacitance and Josephson energy of qubit $q_i$ are denoted by $C_{q_i}$ and $E_J^{q_i}$, respectively, where $i=1,2,3,4$ labels the four qubits. Each qubit is capacitively coupled to two router port couplers $C_{i\mu}$. The self-capacitance and Josephson energy of each port coupler are denoted by $C_{c_{i\mu}}$ and $E_J^{c_{i\mu}}$, respectively, and the qubit--coupler capacitance is $C_{q_i,c_{i\mu}}$. Each port coupler $c_{i\mu}$ is further capacitively coupled to the corresponding bus resonator $m_\mu$, with $\mu=A,B$, through a capacitance $C_{c_{i\mu},m_\mu}$. The self-capacitance and inductive energy of the bus resonator are denoted by $C_{m_\mu}$ and $E_L^{m_\mu}$, respectively. The direct capacitive coupling between each qubit and bus resonator is represented by $C_{q_i,m_\mu}$. The node fluxes are written as $\phi_\lambda$, with $\lambda \in \{q_i, c_{i\mu}, m_\mu\}$. 

The system Lagrangian is
\begin{equation}
L=T-U ,
\end{equation}
where the kinetic energy is
\begin{equation}
\begin{split}
T ={}& \sum_{i=1}^{4}\sum_{\mu=A,B}\frac{1}{2}\Bigl[
 C_{q_i}\dot{\phi}_{q_i}^{2} 
+ C_{c_{i\mu}}\dot{\phi}_{c_{i\mu}}^{2} + C_{m_\mu}\dot{\phi}_{m_\mu}^2\\
& + C_{q_i,c_{i\mu}}\bigl(\dot{\phi}_{q_i}-\dot{\phi}_{c_{i\mu}}\bigr)^{2} \\&
+ C_{c_{i\mu},m_\mu}\bigl(\dot{\phi}_{c_{i\mu}}-\dot{\phi}_{m_\mu}\bigr)^{2} \\
& + C_{q_i,m_\mu}\bigl(\dot{\phi}_{q_i}-\dot{\phi}_{m_\mu}\bigr)^{2}
\Bigr],
\end{split}
\end{equation}
and the potential energy is
\begin{equation}
\begin{split}
U ={}& \sum_{i=1}^{4}\sum_{\mu=A,B}\Bigl(-E_J^{q_i}\cos\varphi_{q_i}
- E_J^{c_{i\mu}}\cos\varphi_{c_{i\mu}} \\
& \quad + \frac{1}{2}E_L^{m_\mu}\varphi_{m_\mu}^{2}
\Bigr),
\end{split}
\end{equation}
where $\varphi_\lambda = 2\pi\phi_\lambda / \Phi_0$ represents the dimensionless gauge-invariant phase. In matrix form, the Lagrangian can be written as
\begin{equation}
L=\frac{1}{2}\dot{\vec{\phi}}^{\,T}\textbf{C}\dot{\vec{\phi}}-U ,
\end{equation}
where $\textbf{C}$ is the capacitance matrix and $\vec{\phi}$ is the vector of node fluxes.

\subsubsection{Hamiltonian}
\label{Hamiltonian}
The canonical conjugate variable of the node flux is defined as
\begin{equation}
q_\lambda=\frac{\partial L}{\partial \dot{\phi}_\lambda}.
\end{equation}
Using $\vec{q}=\textbf{C}\dot{\vec{\phi}}$, the Hamiltonian of the system is obtained via Legendre transformation as
\begin{equation}
H=\sum_{\lambda}q_\lambda \dot{\phi}_\lambda-L
=\frac{1}{2}\vec{q}^{\,T}\textbf{C}^{-1}\vec{q}+U ,
\end{equation}
where $\textbf{C}^{-1}$ denotes the inverse capacitance matrix. In this architecture, the inter-mode capacitive couplings are smaller than the self-capacitance of each mode and larger than the higher-order stray capacitance. After canonical quantization, replacing canonical variables with operators $[\hat{\phi}_\lambda, \hat{q}_{\lambda'}] = i\hbar\delta_{\lambda\lambda'}$, the system Hamiltonian is given by
\begin{equation}
\begin{aligned}
H=\sum_{i,\mu}\Bigg[
&\left(4E_C^{q_i}\hat{n}_{q_i}^{2}-E_J^{q_i}\cos\hat{\varphi}_{q_i}\right)\\&
+\left(4E_C^{c_{i\mu}}\hat{n}_{c_{i\mu}}^{2}
-E_J^{c_{i\mu}}\cos\hat{\varphi}_{c_{i\mu}}\right) \\
&+\left(4E_C^{m_\mu}\hat{n}_{m_\mu}^{2}
+\frac{1}{2}E_L^{m_\mu}\hat{\varphi}_{m_\mu}^{2}\right) \\
&+8\frac{C_{q_i,c_{i\mu}}}{\sqrt{C_{q_i}C_{c_{i\mu}}}}
\sqrt{E_C^{q_i}E_C^{c_{i\mu}}}\,
\hat{n}_{q_i}\hat{n}_{c_{i\mu}} \\
&+8\frac{C_{c_{i\mu},m_\mu}}{\sqrt{C_{c_{i\mu}}C_{m_\mu}}}
\sqrt{E_C^{c_{i\mu}}E_C^{m_\mu}}\,
\hat{n}_{c_{i\mu}}\hat{n}_{m_\mu} \\
&+8(1+\eta)\frac{C_{q_i,m_\mu}}{\sqrt{C_{q_i}C_{m_\mu}}}
\sqrt{E_C^{q_i}E_C^{m_\mu}}\,
\hat{n}_{q_i}\hat{n}_{m_\mu}
\Bigg],
\end{aligned}
\end{equation}
where $\hat{n}_\lambda = \hat{q}_\lambda / (2e)$ is the Cooper-pair number operator and $E_C^\lambda=e^2/(2C_\lambda)$ is the charging energy of mode $\lambda$.

We define the creation and annihilation operators distinctly for the nonlinear transmon modes ($j \in \{q_i, c_{i\mu}\}$) and the linear bus resonators ($\mu \in \{A, B\}$) as
\begin{equation}
\begin{aligned}
\hat{a}_j
&=\frac{1}{\sqrt{2}}
\left[
\left(\frac{E_J^j}{8E_C^j}\right)^{1/4}
\hat{\varphi}_j
+i\left(\frac{8E_C^j}{E_J^j}\right)^{1/4}
\hat{n}_j
\right],\\
\hat{a}_{m_\mu}
&=\frac{1}{\sqrt{2}}
\left[
\left(\frac{E_L^{m_\mu}}{8E_C^{m_\mu}}\right)^{1/4}
\hat{\varphi}_{m_\mu}
+i\left(\frac{8E_C^{m_\mu}}{E_L^{m_\mu}}\right)^{1/4}
\hat{n}_{m_\mu}
\right].
\end{aligned}
\end{equation}
In the transmon regime, $E_J^j \gg E_C^j$, expanding the cosine potential to fourth order ($\cos\hat{\varphi} \approx 1 - \hat{\varphi}^2/2 + \hat{\varphi}^4/24$) and adopting the rotating-wave approximation (RWA), the Hamiltonian can be recast into the multi-mode Duffing-oscillator form:
\begin{equation}
\begin{aligned}
H=&\sum_{i=1}^{4}
\left(
\omega_{q_i}\hat{a}_{q_i}^\dagger\hat{a}_{q_i}
+\frac{\alpha_{q_i}}{2}
\hat{a}_{q_i}^{\dagger 2}\hat{a}_{q_i}^{2}
\right) \\
&+\sum_{i=1}^{4}\sum_{\mu=A,B}
\left(
\omega_{c_{i\mu}}\hat{a}_{c_{i\mu}}^\dagger\hat{a}_{c_{i\mu}}
+\frac{\alpha_{c_{i\mu}}}{2}
\hat{a}_{c_{i\mu}}^{\dagger 2}\hat{a}_{c_{i\mu}}^{2}
\right) \\
&+\sum_{\mu=A,B}
\omega_{m_\mu}\hat{a}_{m_\mu}^\dagger\hat{a}_{m_\mu} \\
&+\sum_{i=1}^{4}\sum_{\mu=A,B}
\Big[
g_{q_i,c_{i\mu}}
\left(\hat{a}_{q_i}^\dagger\hat{a}_{c_{i\mu}}+\mathrm{h.c.}\right)\\&\hspace{2.2cm}
+g_{c_{i\mu},m_\mu}
\left(\hat{a}_{c_{i\mu}}^\dagger\hat{a}_{m_\mu}+\mathrm{h.c.}\right) \\
&\hspace{2.2cm}
+g_{q_i,m_\mu}
\left(\hat{a}_{q_i}^\dagger\hat{a}_{m_\mu}+\mathrm{h.c.}\right)
\Big].
\end{aligned}
\end{equation}
The coupling strengths are determined by the corresponding capacitances as
\begin{equation}
\begin{aligned}
g_{q_i,c_{i\mu}}
&=\frac{1}{2}
\frac{C_{q_i,c_{i\mu}}}{\sqrt{C_{q_i}C_{c_{i\mu}}}}
\sqrt{\omega_{q_i}\omega_{c_{i\mu}}},\\
g_{c_{i\mu},m_\mu}
&=\frac{1}{2}
\frac{C_{c_{i\mu},m_\mu}}{\sqrt{C_{c_{i\mu}}C_{m_\mu}}}
\sqrt{\omega_{c_{i\mu}}\omega_{m_\mu}},\\
g_{q_i,m_\mu}
&=\frac{1+\eta}{2}
\frac{C_{q_i,m_\mu}}{\sqrt{C_{q_i}C_{m_\mu}}}
\sqrt{\omega_{q_i}\omega_{m_\mu}} .
\end{aligned}
\end{equation}
Here $g_{q_i,c_{i\mu}}$, $g_{c_{i\mu},m_\mu}$ and $g_{q_i,m_\mu}$ denote the qubit--coupler, coupler--bus and direct qubit--bus coupling rates, respectively. The dimensionless factor $\eta$ accounts for the correction to the direct qubit--bus capacitive coupling. The qubit and coupler frequencies and anharmonicities are
\begin{equation}
\omega_j=\sqrt{8E_C^j E_J^j}-E_C^j,
\qquad
\alpha_j=-E_C^j, \quad (j \in \{q_i, c_{i\mu}\})
\end{equation}
where $\alpha_j$ is the anharmonicity and the flux-dependent Josephson energy of coupler is given by
\begin{equation}
E_J^{c_{i\mu}}(\Phi_{\mathrm{ext}}) = E_{J0}^{c_{i\mu}} \left| \cos\left( \frac{\pi \Phi_{\mathrm{ext}}}{\Phi_0} \right) \right|,
\label{eq:EJ_flux}
\end{equation}
where $\Phi_0 = h/(2e)$ is the magnetic flux quantum, and $E_{J0}^{c_{i\mu}} = E_{J1}^{c_{i\mu}} + E_{J2}^{c_{i\mu}}$ is the maximum Josephson energy at zero flux. The linear bus-resonator frequency is
\begin{equation}
\omega_{m_\mu}=\sqrt{8E_C^{m_\mu}E_L^{m_\mu}} .
\end{equation}

\subsection{Schrieffer–Wolff transformation}
\label{SW}
In the quantum-router circuit, the couplers are operated in the dispersive regime,
\begin{equation}
\begin{aligned}
\left|\Delta_{q_i,c_{i\mu}}\right|
&= \left|\omega_{q_i}-\omega_{c_{i\mu}}\right|
\gg g_{q_i,c_{i\mu}},\\
\left|\Delta_{m_\mu,c_{i\mu}}\right|
&= \left|\omega_{m_\mu}-\omega_{c_{i\mu}}\right|
\gg g_{c_{i\mu},m_\mu}.
\end{aligned}
\end{equation}
We define the Schrieffer–Wolff generator as
\begin{equation}
\begin{aligned}
S=\sum_{i,\mu}\Bigg[
&\frac{g_{q_i,c_{i\mu}}}{\Delta_{q_i,c_{i\mu}}}
\left(
\hat{a}_{q_i}^{\dagger}\hat{a}_{c_{i\mu}}
-\hat{a}_{c_{i\mu}}^{\dagger}\hat{a}_{q_i}
\right) \\
&+
\frac{g_{c_{i\mu},m_\mu}}{\Delta_{m_\mu,c_{i\mu}}}
\left(
\hat{a}_{m_\mu}^{\dagger}\hat{a}_{c_{i\mu}}
-\hat{a}_{c_{i\mu}}^{\dagger}\hat{a}_{m_\mu}
\right)
\Bigg].
\end{aligned}
\end{equation}
Applying the Schrieffer–Wolff transformation \cite{BRAVYI20112793, PhysRevApplied.10.054062},
\begin{equation}
H_{\mathrm{eff}}=e^{S}He^{-S},
\end{equation}
eliminates the coupler modes to leading order and gives the effective qubit–bus Hamiltonian
\begin{equation}
\begin{aligned}
H_{\mathrm{eff}}
&= \sum_i \tilde{\omega}_{q_i} \hat{a}_{q_i}^{\dagger}\hat{a}_{q_i}
+ \sum_\mu \tilde{\omega}_{m_\mu} \hat{a}_{m_\mu}^{\dagger}\hat{a}_{m_\mu} \\
&\quad + \sum_{i,\mu} \tilde{g}_{q_i,m_\mu}
\left(
\hat{a}_{q_i}^{\dagger}\hat{a}_{m_\mu}
+ \hat{a}_{q_i}\hat{a}_{m_\mu}^{\dagger}
\right),
\end{aligned}
\end{equation}
where the effective qubit–bus coupling is
\begin{equation}
\begin{aligned}
\tilde{g}_{q_i,m_\mu}
&= g_{q_i,m_\mu} + \frac{1}{2} g_{q_i,c_{i\mu}} g_{c_{i\mu},m_\mu} \left(
\frac{1}{\Delta_{q_i,c_{i\mu}}}
+ \frac{1}{\Delta_{m_{\mu},c_{i\mu}}}
\right).
\end{aligned}
\end{equation}
This expression shows that the effective qubit–bus interaction contains both the direct capacitive coupling $g_{q_i,m_\mu}$ and the second-order coupler-mediated contribution.

\subsection{Pulse and Fidelity}
\label{pulse-fidelity}
To switch the effective ZZ interaction on and off, the frequency trajectory of the port coupler \(C_{i\mu}\) is chosen as
\begin{equation}
\begin{split}
\omega_{c_{i\mu}}(t)
&=
\omega_{c_{i\mu}}^{\mathrm{off}}
+
\frac{
\omega_{c_{i\mu}}^{\mathrm{on}}
-
\omega_{c_{i\mu}}^{\mathrm{off}}
}{2}
\biggl[
\operatorname{erf}
\left(
\frac{t-\frac{1}{2}t_{\mathrm{ramp}}}{\sqrt{2}\sigma}
\right) \\
&\quad -
\operatorname{erf}
\left(
\frac{t-t_{\mathrm{gate}}+\frac{1}{2}t_{\mathrm{ramp}}}{\sqrt{2}\sigma}
\right)
\biggr],
\end{split}
\end{equation}
where \(\omega_{c_{i\mu}}^{\mathrm{off}}\) and \(\omega_{c_{i\mu}}^{\mathrm{on}}\) denote the coupler frequencies at the idle and activated operating points, respectively. Here, \(t_{\mathrm{gate}}\) is the total gate duration and \(t_{\mathrm{ramp}}\) is the rise and fall time of the pulse \cite{PhysRevA.87.022309}. This waveform determines the time-dependent Hamiltonian used in the simulation. In the simulation, three energy levels are retained for the qubits and the coupler mode, while two energy levels are retained for the bus resonator.

The gate parameters are optimized by maximizing the average gate fidelity, defined as \cite{PhysRevLett.125.200503}
\begin{equation}
\bar{F}
=
\frac{
\left|\operatorname{Tr}\left(\hat{U}_{\mathrm{id}}^{\dagger}\hat{U}\right)\right|^{2}
+
\operatorname{Tr}\left(\hat{U}^{\dagger}\hat{U}\right)
}
{d(d+1)} ,
\end{equation}
where \(d\) is the dimension of the computational subspace. Thus, \(d=4\) for an isolated two-qubit gate and \(d=16\) for two parallel two-qubit gates. The operator \(\hat{U}\) denotes the evolution operator after single-qubit phase compensation, defined as
\begin{equation}
\hat{U}
=
\left[
\bigotimes_{k=1}^{N}
\hat{R}_{z,k}(\theta_k)
\right]
\hat{U}_{\mathrm{comp}} .
\end{equation}
Here, \(\hat{U}_{\mathrm{comp}}\) is the propagator obtained by solving the time-dependent Schr\"odinger equation for the full system and then projecting the resulting evolution onto the computational subspace. During the gate operation, the qubits can also accumulate single-qubit phases. We therefore apply virtual \(Z\)-gate compensation to each of the \(N\) qubits participating in the gate operation,
\begin{equation}
\hat{R}_{z,k}(\theta_k)
=
\exp\left(-i\boldsymbol{\theta_k}\hat{\sigma}_{z,k}/2\right),
\end{equation}
where the compensation phases \(\boldsymbol{\theta}=\{\theta_1,\ldots,\theta_N\}\) are treated as free parameters and numerically optimized over the interval \([-\pi,\pi]\) to maximize \(\bar{F}\).
In the numerical simulations, the qubits and couplers are truncated to three levels, whereas the coplanar waveguide resonators are truncated to two levels.

\subsection{Hardware topology and compilation}\label{Hardware and topo}
\subsubsection{Hardware Topology}
\label{topo}
We model the modular chip topology as an undirected graph $G=(V,E)$, where $V$ denotes the set of physical qubits and $E$ represents the physical connections supporting native CZ gates. The two architectures share identical intra-module connections, $E_{\text{intra}}$, differing solely in their inter-module connectivity. 

The Grid topology is defined as $G_{\text{grid}} = (V,E_{\text{grid}})$, with $E_{\text{grid}} = E_{\text{intra}} \cup E_{\text{inter}}^{\text{grid}}$. Here, $E_{\text{inter}}^{\text{grid}}$ consists strictly of fixed nearest-neighbour connections between adjacent modules. In contrast, the Router topology augments the Grid connectivity by introducing cross-module connections, $E_{\text{inter}}^{\text{cross}}$. It is formulated as $G_{\text{router}} = (V,E_{\text{router}})$, where $E_{\text{router}} = E_{\text{grid}} \cup E_{\text{inter}}^{\text{cross}}$. Consequently, the relation $E_{\text{grid}} \subseteq E_{\text{router}}$ guarantees that any quantum circuit validly compiled for the Grid topology can be inherently executed on the Router topology.

\subsubsection{SABRE Compilation and Native Gates}
\label{Sabre}
For each circuit class and system scale, the logical circuits are directly compiled onto both the Grid and Router topologies. Both architectures are evaluated under identical Qiskit compilation settings, utilizing the SABRE algorithm for initial layout and routing, optimization level 3, and a consistent set of 100 random seeds ($s = 0, \dots, 99$).

The initial compilation targets the basis gate set $\{R_z, \sqrt{X}, X, \text{CZ}, \text{SWAP}\}$ to explicitly retain the SWAP gates inserted by the compiler. Subsequently, the circuits are further decomposed into the hardware-native gate set $\{R_z, \sqrt{X}, X, \text{CZ}\}$. This procedure yields two key gate-count metrics: $N_{\text{SWAP}}$, representing the number of SWAP gates introduced to satisfy hardware connectivity constraints, and $N_{\text{CZ}}$, denoting the total number of CZ gates post-decomposition. Specifically, $N_{\text{CZ}}$ encompasses both the intrinsic CZ gates originally required by the algorithm and the supplementary CZ gates generated from the SWAP decompositions.

\subsubsection{Router achievable envelope}
\label{router envelope}
For each random seed $s$, we generate two candidate circuits: $C_{\text{direct}, s}$, compiled directly onto the Router topology, and $C_{\text{grid}, s}$, compiled onto the Grid topology. Because the Grid represents a subgraph of the Router, the circuit $C_{\text{grid}, s}$ is inherently executable on the Router architecture. Consequently, we define the minimum gate counts for both SWAP and CZ metrics between these two candidate circuits as:
\begin{equation}
\begin{aligned}
N_{\text{SWAP}}^{\text{env}}(s) &= \min \left[ N_{\text{SWAP}}^{\text{grid}}(s), N_{\text{SWAP}}^{\text{direct}}(s) \right], \\
N_{\text{CZ}}^{\text{env}}(s) &= \min \left[ N_{\text{CZ}}^{\text{grid}}(s), N_{\text{CZ}}^{\text{direct}}(s) \right].
\end{aligned}
\end{equation}
We refer to these quantities as the \textit{Router achievable envelope}. This envelope objectively reflects the optimal gate-count performance attainable on the Router architecture by dynamically adopting the superior outcome from the two compilation strategies.

\subsubsection{Serial and Parallel Scheduling Depths}
\label{deepth}
The depth of the native circuit is evaluated using an as-soon-as-possible (ASAP) scheduling strategy. Under this protocol, each gate is executed immediately once its preceding operational dependencies are resolved and the requisite hardware resources become available. The model assumes a uniform unit-time duration for all native gates and strictly dictates that gates acting on overlapping qubits cannot be assigned to the same scheduling layer.

For the Router architecture, we investigate two distinct resource allocation schemes. The \textit{serial} mode restricts a given module boundary to mediating at most one cross-module CZ gate per scheduling layer. Conversely, the \textit{parallel} mode doubles this capacity, permitting the parallel execution of two cross-module CZ gates across the same boundary, provided they act on disjoint qubit pairs. Note that routing resources across different module boundaries operate entirely independently.

For a given algorithm scale and SABRE seed, the optimal candidate circuit $C_s^*$ is selected from $\{C_{\text{direct},s}, C_{\text{grid},s}\}$. The primary selection criterion is the minimization of circuit depth. In the event of a depth tie, the algorithm sequentially assesses the CZ gate count and the SWAP gate count. This lexicographical minimization is formally expressed as:
\begin{equation}
\begin{aligned}
C_s^* = {} & \underset{C \in \{C_{\text{direct},s}, C_{\text{grid},s}\}}{\operatorname{arg\,min}_{\text{lex}}} \\
& \Big[ D_{\text{parallel}}(C), N_{\text{CZ}}(C), N_{\text{SWAP}}(C) \Big],
\end{aligned}
\end{equation}
where the subscript ``lex'' denotes an item-by-item comparison in the specified order. Once the optimal circuit $C_s^*$ is determined, both the serial and parallel routing capacity constraints are applied to this exact same circuit independently:
\begin{equation}
D_{\text{router}}^{\text{serial}} = D_{\text{serial}}(C_s^*), \quad D_{\text{router}}^{\text{parallel}} = D_{\text{parallel}}(C_s^*).
\end{equation}
Consequently, the comparison between serial and parallel depths solely modifies the available capacity at the router boundaries, keeping the logical circuit layout, gate counts, operational sequence, and qubit mapping strictly invariant. This rigorous control ensures that the depth reduction, $D_{\text{router}}^{\text{serial}} - D_{\text{router}}^{\text{parallel}}$, unambiguously reflects the parallel scheduling benefit afforded by the routing resources, explicitly excluding any artifacts stemming from variations between different compiled circuits. As a baseline reference, $D_{\text{grid}}(s)$ represents the depth of the Grid-compiled circuit evaluated under standard ASAP scheduling.

\subsubsection{Statistical Methods and Performance Metrics}
\label{statistical method}
For each circuit class and system scale, the solid lines plotted in the figures represent the median values obtained across the 100 random seeds. To rigorously evaluate the performance gains, the relative improvements are computed on a per-seed basis as follows:
\begin{equation}
\begin{aligned}
\Delta_{\text{SWAP}}(s) &= \frac{N_{\text{SWAP}}^{\text{grid}}(s) - N_{\text{SWAP}}^{\text{env}}(s)}{N_{\text{SWAP}}^{\text{grid}}(s)}, \\
\Delta_{\text{CZ}}(s) &= \frac{N_{\text{CZ}}^{\text{grid}}(s) - N_{\text{CZ}}^{\text{env}}(s)}{N_{\text{CZ}}^{\text{grid}}(s)}, \\
\Delta_{D_\text{parallel}}(s) &= \frac{D_{\text{serial}}(s) - D_{\text{parallel}}(s)}{D_{\text{serial}}(s)}, \\
\Delta_{D_\text{total}}(s) &= \frac{D_{\text{grid}}(s) - D_{\text{parallel}}(s)}{D_{\text{grid}}(s)}.
\end{aligned}
\end{equation}
Here, $\Delta_{D_\text{parallel}}$ quantifies the relative depth reduction achieved by the parallel mode over the serial mode, directly isolating the scheduling advantage. Similarly, $\Delta_{D_\text{total}}$ evaluates the comprehensive, end-to-end depth improvement of the parallel Router architecture compared to the baseline Grid topology.

\section*{Declarations}

\phantomsection
\bmhead{Acknowledgements}

The authors would like to thank Zhaoqiu Yang, Zican Dong, and Hang Lian for their helpful discussions on compiler-related topics. This work was supported by the National Key Research and Development Program of China (Grant No. 2024YFB4504101) and the National Natural Science Foundation of China (Grant No. 62636009).

\begin{itemize}

\item Funding 

This work was supported by the National Key Research and Development Program of China (Grant No. 2024YFB4504101) and the National Natural Science Foundation of China (Grant No. 62636009).

\item Conflict of interest/Competing interests 

The authors declare no competing interests.

\item Ethics approval and consent to participate

Not applicable
\item Consent for publication

Not applicable
\item Data availability

The data that support the findings of this study are available from the corresponding author upon reasonable request.

\item Code availability 

Codes are available from the corresponding author upon reasonable request.

\item Author contribution

\bmhead{Author contributions}
Z.S. and W.W. conceived and supervised the project. B.Y. developed the theoretical model, conducted circuit-level simulations, and analyzed the data. C.Z. and H.H. assisted with the theoretical derivations and simulation configurations. Y.F. and C.H. contributed to parameter optimization and data validation. H.S., B.Z., and F.L. contributed to discussions on hardware implementation and architectural scalability. B.Y. wrote the manuscript with critical input from W.W. and Z.S. All authors discussed the results and approved the final manuscript.

\end{itemize}

\bibliography{sn-bibliography}% common bib file

@article{williamson2026low,
  title={Low-overhead fault-tolerant quantum computation by gauging logical operators},
  author={Williamson, Dominic J and Yoder, Theodore J},
  journal={Nature Physics},
  pages={1--6},
  year={2026},
  publisher={Nature Publishing Group UK London}
}

@INPROCEEDINGS{shor,
  author={Shor, P.W.},
  booktitle={Proceedings of 37th Conference on Foundations of Computer Science}, 
  title={Fault-tolerant quantum computation}, 
  year={1996},
  volume={},
  number={},
  pages={56-65},
  doi={10.1109/SFCS.1996.548464}}

@article{krinner2022realizing,
  title={Realizing repeated quantum error correction in a distance-three surface code},
  author={Krinner, Sebastian and Lacroix, Nathan and Remm, Ants and Di Paolo, Agustin and Genois, Elie and Leroux, Catherine and Hellings, Christoph and Lazar, Stefania and Swiadek, Francois and Herrmann, Johannes and others},
  journal={Nature},
  volume={605},
  number={7911},
  pages={669--674},
  year={2022},
  publisher={Nature Publishing Group UK London}
}

@article{google2025quantum,
  title={Quantum error correction below the surface code threshold},
  journal={Nature},
  volume={638},
  number={8052},
  pages={920--926},
  year={2025},
  publisher={Nature Publishing Group UK London}
}

@article{Zhou2025,
author = {Zhou, Hengyun and Zhao, Chen and Cain, Madelyn and Bluvstein, Dolev and Maskara, Nishad and Duckering, Casey and Hu, Hong-Ye and Wang, Sheng-Tao and Kubica, Aleksander and Lukin, Mikhail D},
doi = {10.1038/s41586-025-09543-5},
issn = {1476-4687},
journal = {Nature},
number = {8084},
pages = {303--308},
title = {{Low-overhead transversal fault tolerance for universal quantum computation}},
url = {https://doi.org/10.1038/s41586-025-09543-5},
volume = {646},
year = {2025}
}

@article{Acharya2023,
author = {Acharya, Rajeev and Aleiner, Igor and Allen, Richard and Andersen, Trond I and Ansmann, Markus and Arute, Frank and Arya, Kunal and Asfaw, Abraham and Atalaya, Juan and Babbush, Ryan and Bacon, Dave and Bardin, Joseph C and Basso, Joao and Bengtsson, Andreas and Boixo, Sergio and Bortoli, Gina and Bourassa, Alexandre and Bovaird, Jenna and Brill, Leon and Broughton, Michael and Buckley, Bob B and Buell, David A and Burger, Tim and Burkett, Brian and Bushnell, Nicholas and Chen, Yu and Chen, Zijun and Chiaro, Ben and Cogan, Josh and Collins, Roberto and Conner, Paul and Courtney, William and Crook, Alexander L and Curtin, Ben and Debroy, Dripto M and {Del Toro Barba}, Alexander and Demura, Sean and Dunsworth, Andrew and Eppens, Daniel and Erickson, Catherine and Faoro, Lara and Farhi, Edward and Fatemi, Reza and {Flores Burgos}, Leslie and Forati, Ebrahim and Fowler, Austin G and Foxen, Brooks and Giang, William and Gidney, Craig and Gilboa, Dar and Giustina, Marissa and {Grajales Dau}, Alejandro and Gross, Jonathan A and Habegger, Steve and Hamilton, Michael C and Harrigan, Matthew P and Harrington, Sean D and Higgott, Oscar and Hilton, Jeremy and Hoffmann, Markus and Hong, Sabrina and Huang, Trent and Huff, Ashley and Huggins, William J and Ioffe, Lev B and Isakov, Sergei V and Iveland, Justin and Jeffrey, Evan and Jiang, Zhang and Jones, Cody and Juhas, Pavol and Kafri, Dvir and Kechedzhi, Kostyantyn and Kelly, Julian and Khattar, Tanuj and Khezri, Mostafa and Kieferov{\'{a}}, M{\'{a}}ria and Kim, Seon and Kitaev, Alexei and Klimov, Paul V and Klots, Andrey R and Korotkov, Alexander N and Kostritsa, Fedor and Kreikebaum, John Mark and Landhuis, David and Laptev, Pavel and Lau, Kim-Ming and Laws, Lily and Lee, Joonho and Lee, Kenny and Lester, Brian J and Lill, Alexander and Liu, Wayne and Locharla, Aditya and Lucero, Erik and Malone, Fionn D and Marshall, Jeffrey and Martin, Orion and McClean, Jarrod R and McCourt, Trevor and McEwen, Matt and Megrant, Anthony and {Meurer Costa}, Bernardo and Mi, Xiao and Miao, Kevin C and Mohseni, Masoud and Montazeri, Shirin and Morvan, Alexis and Mount, Emily and Mruczkiewicz, Wojciech and Naaman, Ofer and Neeley, Matthew and Neill, Charles and Nersisyan, Ani and Neven, Hartmut and Newman, Michael and Ng, Jiun How and Nguyen, Anthony and Nguyen, Murray and Niu, Murphy Yuezhen and O'Brien, Thomas E and Opremcak, Alex and Platt, John and Petukhov, Andre and Potter, Rebecca and Pryadko, Leonid P and Quintana, Chris and Roushan, Pedram and Rubin, Nicholas C and Saei, Negar and Sank, Daniel and Sankaragomathi, Kannan and Satzinger, Kevin J and Schurkus, Henry F and Schuster, Christopher and Shearn, Michael J and Shorter, Aaron and Shvarts, Vladimir and Skruzny, Jindra and Smelyanskiy, Vadim and Smith, W Clarke and Sterling, George and Strain, Doug and Szalay, Marco and Torres, Alfredo and Vidal, Guifre and Villalonga, Benjamin and {Vollgraff Heidweiller}, Catherine and White, Theodore and Xing, Cheng and Yao, Z Jamie and Yeh, Ping and Yoo, Juhwan and Young, Grayson and Zalcman, Adam and Zhang, Yaxing and Zhu, Ningfeng and AI, Google Quantum},
doi = {10.1038/s41586-022-05434-1},
issn = {1476-4687},
journal = {Nature},
number = {7949},
pages = {676--681},
title = {{Suppressing quantum errors by scaling a surface code logical qubit}},
url = {https://doi.org/10.1038/s41586-022-05434-1},
volume = {614},
year = {2023}
}

@article{whhNc,
Author = {Zhang, Aosai and Xie, Haipeng and Gao, Yu and Yang, Jia-Nan and Bao,
   Zehang and Zhu, Zitian and Chen, Jiachen and Wang, Ning and Zhang,
   Chuanyu and Zhong, Jiarun and Xu, Shibo and Wang, Ke and Wu, Yaozu and
   Jin, Feitong and Zhu, Xuhao and Zou, Yiren and Tan, Ziqi and Cui,
   Zhengyi and Shen, Fanhao and Li, Tingting and Han, Yihang and He, Yiyang
   and Liu, Gongyu and Shen, Jiayuan and Wang, Han and Wang, Yanzhe and
   Dong, Hang and Deng, Jinfeng and Li, Hekang and Wang, Zhen and Song,
   Chao and Guo, Qiujiang and Zhang, Pengfei and Li, Ying and Wang, H.},
Title = {Demonstrating quantum error mitigation on logical qubits},
Journal = {NATURE COMMUNICATIONS},
Year = {2025},
Volume = {17},
Number = {1},
Month = {DEC 24},
DOI = {10.1038/s41467-025-67768-4},
Article-Number = {1021},
EISSN = {2041-1723},
ResearcherID-Numbers = {deng, jinfeng/QSP-0986-2026
   Guo, Qiujiang/HCI-0720-2022
   Zhang, Pengfei/PJS-8509-2026
   Bao, Zehang/OIT-9376-2025
   Wang, Haohua/G-8875-2011
   Li, Ying/I-1302-2013
   Tan, Ziqi/GRR-6050-2022},
Unique-ID = {WOS:001672449700003},
}

@article{siddiqi2021engineering,
  title={Engineering high-coherence superconducting qubits},
  author={Siddiqi, Irfan},
  journal={Nature Reviews Materials},
  volume={6},
  number={10},
  pages={875--891},
  year={2021},
  publisher={Nature Publishing Group UK London}
}

@article{dd96-gcb6,
  title = {Nondegenerate Noise-Resilient Superconducting Qubit},
  author = {Hays, Max and Kim, Junghyun and Oliver, William D.},
  journal = {PRX Quantum},
  volume = {6},
  issue = {4},
  pages = {040321},
  numpages = {19},
  year = {2025},
  month = {Oct},
  publisher = {American Physical Society},
  doi = {10.1103/dd96-gcb6},
  url = {https://link.aps.org/doi/10.1103/dd96-gcb6}
}

@article{ganjam2024surpassing,
  title={Surpassing millisecond coherence in on chip superconducting quantum memories by optimizing materials and circuit design},
  author={Ganjam, Suhas and Wang, Yanhao and Lu, Yao and Banerjee, Archan and Lei, Chan U and Krayzman, Lev and Kisslinger, Kim and Zhou, Chenyu and Li, Ruoshui and Jia, Yichen and others},
  journal={Nature Communications},
  volume={15},
  number={1},
  pages={3687},
  year={2024},
  publisher={Nature Publishing Group UK London}
}

@article{Gao2025,
author = {Gao, Ran and Wu, Feng and Sun, Hantao and Chen, Jianjun and Deng, Hao and Ma, Xizheng and Miao, Xiaohe and Song, Zhijun and Wan, Xin and Wang, Fei and Xia, Tian and Ying, Make and Zhang, Chao and Shi, Yaoyun and Zhao, Hui-Hai and Deng, Chunqing},
doi = {10.1038/s41467-025-58745-y},
issn = {2041-1723},
journal = {Nature Communications},
number = {1},
pages = {3620},
title = {{The effects of disorder in superconducting materials on qubit coherence}},
url = {https://doi.org/10.1038/s41467-025-58745-y},
volume = {16},
year = {2025}
}

@article{Tuokkola2025,
author = {Tuokkola, Mikko and Sunada, Yoshiki and Kivij{\"{a}}rvi, Heidi and Albanese, Jonatan and Gr{\"{o}}nberg, Leif and Kaikkonen, Jukka-Pekka and Vesterinen, Visa and Govenius, Joonas and M{\"{o}}tt{\"{o}}nen, Mikko},
doi = {10.1038/s41467-025-61126-0},
issn = {2041-1723},
journal = {Nature Communications},
number = {1},
pages = {5421},
title = {{Methods to achieve near-millisecond energy relaxation and dephasing times for a superconducting transmon qubit}},
url = {https://doi.org/10.1038/s41467-025-61126-0},
volume = {16},
year = {2025}
}

@article{Bland2025,
author = {Bland, Matthew P and Bahrami, Faranak and Martinez, Jeronimo G C and Prestegaard, Paal H and Smitham, Basil M and Joshi, Atharv and Hedrick, Elizabeth and Kumar, Shashwat and Yang, Ambrose and Pakpour-Tabrizi, Alexander C and Jindal, Apoorv and Chang, Ray D and Cheng, Guangming and Yao, Nan and Cava, Robert J and de Leon, Nathalie P and Houck, Andrew A},
doi = {10.1038/s41586-025-09687-4},
issn = {1476-4687},
journal = {Nature},
number = {8089},
pages = {343--348},
title = {{Millisecond lifetimes and coherence times in 2D transmon qubits}},
url = {https://doi.org/10.1038/s41586-025-09687-4},
volume = {647},
year = {2025}
}

@article{Dane2026,
author = {Dane, Andrew and Balakrishnan, Karthik and Wacaser, Brent and Hung, Li-Wen and Mamin, H J and Rugar, Daniel and Shelby, Robert M and Murray, Conal and Rodbell, Kenneth and Sleight, Jeffrey},
doi = {10.1038/s41534-026-01199-x},
issn = {2056-6387},
journal = {npj Quantum Information},
number = {1},
pages = {62},
title = {{Robust quality factor assessment of high-coherence superconducting qubits}},
url = {https://doi.org/10.1038/s41534-026-01199-x},
volume = {12},
year = {2026}
}

@article{Smith2020,
author = {Smith, W C and Kou, A and Xiao, X and Vool, U and Devoret, M H},
doi = {10.1038/s41534-019-0231-2},
issn = {2056-6387},
journal = {npj Quantum Information},
number = {1},
pages = {8},
title = {{Superconducting circuit protected by two-Cooper-pair tunneling}},
url = {https://doi.org/10.1038/s41534-019-0231-2},
volume = {6},
year = {2020}
}

@article{Li2023,
author = {Li, Zhiyuan and Liu, Pei and Zhao, Peng and Mi, Zhenyu and Xu, Huikai and Liang, Xuehui and Su, Tang and Sun, Weijie and Xue, Guangming and Zhang, Jing-Ning and Liu, Weiyang and Jin, Yirong and Yu, Haifeng},
doi = {10.1038/s41534-023-00781-x},
issn = {2056-6387},
journal = {npj Quantum Information},
number = {1},
pages = {111},
title = {{Error per single-qubit gate below 10−4 in a superconducting qubit}},
url = {https://doi.org/10.1038/s41534-023-00781-x},
volume = {9},
year = {2023}
}

@article{Li2019,
author = {Li, Shaowei and Clark, Juno and Wang, Shiyu and Wu, Yulin and Gong, Ming and Yan, Zhiguang and Rong, Hao and Deng, Hui and Zha, Chen and Guo, Cheng and Sun, Lihua and Peng, Chengzhi and Zhu, Xiaobo and Pan, Jian-Wei},
doi = {10.1038/s41534-019-0202-7},
issn = {2056-6387},
journal = {npj Quantum Information},
number = {1},
pages = {84},
title = {{Realisation of high-fidelity nonadiabatic CZ gates with superconducting qubits}},
url = {https://doi.org/10.1038/s41534-019-0202-7},
volume = {5},
year = {2019}
}

@article{PhysRevX.11.021058,
  title = {Realization of High-Fidelity CZ and $ZZ$-Free iSWAP Gates with a Tunable Coupler},
  author = {Sung, Youngkyu and Ding, Leon and Braum\"uller, Jochen and Veps\"al\"ainen, Antti and Kannan, Bharath and Kjaergaard, Morten and Greene, Ami and Samach, Gabriel O. and McNally, Chris and Kim, David and Melville, Alexander and Niedzielski, Bethany M. and Schwartz, Mollie E. and Yoder, Jonilyn L. and Orlando, Terry P. and Gustavsson, Simon and Oliver, William D.},
  journal = {Phys. Rev. X},
  volume = {11},
  issue = {2},
  pages = {021058},
  numpages = {32},
  year = {2021},
  month = {Jun},
  publisher = {American Physical Society},
  doi = {10.1103/PhysRevX.11.021058},
  url = {https://link.aps.org/doi/10.1103/PhysRevX.11.021058}
}

@article{rf7g-md41,
  title = {Crosstalk in multiqubit fluxonium architectures with transmon couplers},
  author = {Zwanenburg, Martijn F. S. and Andersen, Christian Kraglund},
  journal = {Phys. Rev. Appl.},
  volume = {26},
  issue = {2},
  pages = {024088},
  numpages = {14},
  year = {2026},
  month = {Aug},
  publisher = {American Physical Society},
  doi = {10.1103/rf7g-md41},
  url = {https://link.aps.org/doi/10.1103/rf7g-md41}
}

@article{lvb9-pfr3,
  title = {Direct Implementation of High-Fidelity Three-Qubit Gates for Superconducting Processor with Tunable Couplers},
  author = {Liu, Hao-Tian and Chen, Bing-Jie and Zhang, Jia-Chi and Xiao, Yong-Xi and Li, Tian-Ming and Huang, Kaixuan and Wang, Ziting and Li, Hao and Zhao, Kui and Xu, Yueshan and Deng, Cheng-Lin and Liang, Gui-Han and Liu, Zheng-He and Zhou, Si-Yun and Fang, Cai-Ping and Song, Xiaohui and Xiang, Zhongcheng and Zheng, Dongning and Shi, Yun-Hao and Xu, Kai and Fan, Heng},
  journal = {Phys. Rev. Lett.},
  volume = {135},
  issue = {5},
  pages = {050602},
  numpages = {8},
  year = {2025},
  month = {Jul},
  publisher = {American Physical Society},
  doi = {10.1103/lvb9-pfr3},
  url = {https://link.aps.org/doi/10.1103/lvb9-pfr3}
}

@article{PhysRevLett.125.240503,
  title = {High-Fidelity, High-Scalability Two-Qubit Gate Scheme for Superconducting Qubits},
  author = {Xu, Yuan and Chu, Ji and Yuan, Jiahao and Qiu, Jiawei and Zhou, Yuxuan and Zhang, Libo and Tan, Xinsheng and Yu, Yang and Liu, Song and Li, Jian and Yan, Fei and Yu, Dapeng},
  journal = {Phys. Rev. Lett.},
  volume = {125},
  issue = {24},
  pages = {240503},
  numpages = {7},
  year = {2020},
  month = {Dec},
  publisher = {American Physical Society},
  doi = {10.1103/PhysRevLett.125.240503},
  url = {https://link.aps.org/doi/10.1103/PhysRevLett.125.240503}
}

@article{PhysRevLett.129.060501,
  title = {Hamiltonian Engineering with Multicolor Drives for Fast Entangling Gates and Quantum Crosstalk Cancellation},
  author = {Wei, K. X. and Magesan, E. and Lauer, I. and Srinivasan, S. and Bogorin, D. F. and Carnevale, S. and Keefe, G. A. and Kim, Y. and Klaus, D. and Landers, W. and Sundaresan, N. and Wang, C. and Zhang, E. J. and Steffen, M. and Dial, O. E. and McKay, D. C. and Kandala, A.},
  journal = {Phys. Rev. Lett.},
  volume = {129},
  issue = {6},
  pages = {060501},
  numpages = {6},
  year = {2022},
  month = {Aug},
  publisher = {American Physical Society},
  doi = {10.1103/PhysRevLett.129.060501},
  url = {https://link.aps.org/doi/10.1103/PhysRevLett.129.060501}
}

@article{Kim2022,
author = {Kim, Yosep and Morvan, Alexis and Nguyen, Long B and Naik, Ravi K and J{\"{u}}nger, Christian and Chen, Larry and Kreikebaum, John Mark and Santiago, David I and Siddiqi, Irfan},
doi = {10.1038/s41567-022-01590-3},
issn = {1745-2481},
journal = {Nature Physics},
number = {7},
pages = {783--788},
title = {{High-fidelity three-qubit iToffoli gate for fixed-frequency superconducting qubits}},
url = {https://doi.org/10.1038/s41567-022-01590-3},
volume = {18},
year = {2022}
}

@article{arute2019quantum,
  title={Quantum supremacy using a programmable superconducting processor},
  author={Arute, Frank and Arya, Kunal and Babbush, Ryan and Bacon, Dave and Bardin, Joseph C and Barends, Rami and Biswas, Rupak and Boixo, Sergio and Brandao, Fernando GSL and Buell, David A and others},
  journal={nature},
  volume={574},
  number={7779},
  pages={505--510},
  year={2019},
  publisher={Nature Publishing Group UK London}
}

@article{PhysRevLett.127.180501,
  title = {Strong Quantum Computational Advantage Using a Superconducting Quantum Processor},
  author = {Wu, Yulin and Bao, Wan-Su and Cao, Sirui and Chen, Fusheng and Chen, Ming-Cheng and Chen, Xiawei and Chung, Tung-Hsun and Deng, Hui and Du, Yajie and Fan, Daojin and Gong, Ming and Guo, Cheng and Guo, Chu and Guo, Shaojun and Han, Lianchen and Hong, Linyin and Huang, He-Liang and Huo, Yong-Heng and Li, Liping and Li, Na and Li, Shaowei and Li, Yuan and Liang, Futian and Lin, Chun and Lin, Jin and Qian, Haoran and Qiao, Dan and Rong, Hao and Su, Hong and Sun, Lihua and Wang, Liangyuan and Wang, Shiyu and Wu, Dachao and Xu, Yu and Yan, Kai and Yang, Weifeng and Yang, Yang and Ye, Yangsen and Yin, Jianghan and Ying, Chong and Yu, Jiale and Zha, Chen and Zhang, Cha and Zhang, Haibin and Zhang, Kaili and Zhang, Yiming and Zhao, Han and Zhao, Youwei and Zhou, Liang and Zhu, Qingling and Lu, Chao-Yang and Peng, Cheng-Zhi and Zhu, Xiaobo and Pan, Jian-Wei},
  journal = {Phys. Rev. Lett.},
  volume = {127},
  issue = {18},
  pages = {180501},
  numpages = {7},
  year = {2021},
  month = {Oct},
  publisher = {American Physical Society},
  doi = {10.1103/PhysRevLett.127.180501},
  url = {https://link.aps.org/doi/10.1103/PhysRevLett.127.180501}
}

@article{Kim2023,
author = {Kim, Youngseok and Eddins, Andrew and Anand, Sajant and Wei, Ken Xuan and van den Berg, Ewout and Rosenblatt, Sami and Nayfeh, Hasan and Wu, Yantao and Zaletel, Michael and Temme, Kristan and Kandala, Abhinav},
doi = {10.1038/s41586-023-06096-3},
issn = {1476-4687},
journal = {Nature},
number = {7965},
pages = {500--505},
title = {{Evidence for the utility of quantum computing before fault tolerance}},
url = {https://doi.org/10.1038/s41586-023-06096-3},
volume = {618},
year = {2023}
}

@article{PhysRevLett.134.090601,
  title = {Establishing a New Benchmark in Quantum Computational Advantage with 105-qubit Zuchongzhi 3.0 Processor},
  author = {Gao, Dongxin and Fan, Daojin and Zha, Chen and Bei, Jiahao and Cai, Guoqing and Cai, Jianbin and Cao, Sirui and Chen, Fusheng and Chen, Jiang and Chen, Kefu and Chen, Xiawei and Chen, Xiqing and Chen, Zhe and Chen, Zhiyuan and Chen, Zihua and Chu, Wenhao and Deng, Hui and Deng, Zhibin and Ding, Pei and Ding, Xun and Ding, Zhuzhengqi and Dong, Shuai and Dong, Yupeng and Fan, Bo and Fu, Yuanhao and Gao, Song and Ge, Lei and Gong, Ming and Gui, Jiacheng and Guo, Cheng and Guo, Shaojun and Guo, Xiaoyang and Han, Lianchen and He, Tan and Hong, Linyin and Hu, Yisen and Huang, He-Liang and Huo, Yong-Heng and Jiang, Tao and Jiang, Zuokai and Jin, Honghong and Leng, Yunxiang and Li, Dayu and Li, Dongdong and Li, Fangyu and Li, Jiaqi and Li, Jinjin and Li, Junyan and Li, Junyun and Li, Na and Li, Shaowei and Li, Wei and Li, Yuhuai and Li, Yuan and Liang, Futian and Liang, Xuelian and Liao, Nanxing and Lin, Jin and Lin, Weiping and Liu, Dailin and Liu, Hongxiu and Liu, Maliang and Liu, Xinyu and Liu, Xuemeng and Liu, Yancheng and Lou, Haoxin and Ma, Yuwei and Meng, Lingxin and Mou, Hao and Nan, Kailiang and Nie, Binghan and Nie, Meijuan and Ning, Jie and Niu, Le and Peng, Wenyi and Qian, Haoran and Rong, Hao and Rong, Tao and Shen, Huiyan and Shen, Qiong and Su, Hong and Su, Feifan and Sun, Chenyin and Sun, Liangchao and Sun, Tianzuo and Sun, Yingxiu and Tan, Yimeng and Tan, Jun and Tang, Longyue and Tu, Wenbing and Wan, Cai and Wang, Jiafei and Wang, Biao and Wang, Chang and Wang, Chen and Wang, Chu and Wang, Jian and Wang, Liangyuan and Wang, Rui and Wang, Shengtao and Wang, Xiaomin and Wang, Xinzhe and Wang, Xunxun and Wang, Yeru and Wei, Zuolin and Wei, Jiazhou and Wu, Dachao and Wu, Gang and Wu, Jin and Wu, Shengjie and Wu, Yulin and Xie, Shiyong and Xin, Lianjie and Xu, Yu and Xue, Chun and Yan, Kai and Yang, Weifeng and Yang, Xinpeng and Yang, Yang and Ye, Yangsen and Ye, Zhenping and Ying, Chong and Yu, Jiale and Yu, Qinjing and Yu, Wenhu and Zeng, Xiangdong and Zhan, Shaoyu and Zhang, Feifei and Zhang, Haibin and Zhang, Kaili and Zhang, Pan and Zhang, Wen and Zhang, Yiming and Zhang, Yongzhuo and Zhang, Lixiang and Zhao, Guming and Zhao, Peng and Zhao, Xianhe and Zhao, Xintao and Zhao, Youwei and Zhao, Zhong and Zheng, Luyuan and Zhou, Fei and Zhou, Liang and Zhou, Na and Zhou, Naibin and Zhou, Shifeng and Zhou, Shuang and Zhou, Zhengxiao and Zhu, Chengjun and Zhu, Qingling and Zou, Guihong and Zou, Haonan and Zhang, Qiang and Lu, Chao-Yang and Peng, Cheng-Zhi and Zhu, Xiaobo and Pan, Jian-Wei},
  journal = {Phys. Rev. Lett.},
  volume = {134},
  issue = {9},
  pages = {090601},
  numpages = {7},
  year = {2025},
  month = {Mar},
  publisher = {American Physical Society},
  doi = {10.1103/PhysRevLett.134.090601},
  url = {https://link.aps.org/doi/10.1103/PhysRevLett.134.090601}
}

@article{Niu2023,
author = {Niu, Jingjing and Zhang, Libo and Liu, Yang and Qiu, Jiawei and Huang, Wenhui and Huang, Jiaxiang and Jia, Hao and Liu, Jiawei and Tao, Ziyu and Wei, Weiwei and Zhou, Yuxuan and Zou, Wanjing and Chen, Yuanzhen and Deng, Xiaowei and Deng, Xiuhao and Hu, Changkang and Hu, Ling and Li, Jian and Tan, Dian and Xu, Yuan and Yan, Fei and Yan, Tongxing and Liu, Song and Zhong, Youpeng and Cleland, Andrew N and Yu, Dapeng},
doi = {10.1038/s41928-023-00925-z},
issn = {2520-1131},
journal = {Nature Electronics},
number = {3},
pages = {235--241},
title = {{Low-loss interconnects for modular superconducting quantum processors}},
url = {https://doi.org/10.1038/s41928-023-00925-z},
volume = {6},
year = {2023}
}

@article{Gold2021,
author = {Gold, Alysson and Paquette, J P and Stockklauser, Anna and Reagor, Matthew J and Alam, M Sohaib and Bestwick, Andrew and Didier, Nicolas and Nersisyan, Ani and Oruc, Feyza and Razavi, Armin and Scharmann, Ben and Sete, Eyob A and Sur, Biswajit and Venturelli, Davide and Winkleblack, Cody James and Wudarski, Filip and Harburn, Mike and Rigetti, Chad},
doi = {10.1038/s41534-021-00484-1},
issn = {2056-6387},
journal = {npj Quantum Information},
number = {1},
pages = {142},
title = {{Entanglement across separate silicon dies in a modular superconducting qubit device}},
url = {https://doi.org/10.1038/s41534-021-00484-1},
volume = {7},
year = {2021}
}

@article{Vezvaee2026,
author = {Vezvaee, Arian and Benito, Cesar and Morford-Oberst, Mario and Bermudez, Alejandro and Lidar, Daniel A},
doi = {10.1038/s41467-026-76090-6},
issn = {2041-1723},
journal = {Nature Communications},
number = {1},
pages = {9201},
title = {{Surface code scaling on heavy-hex superconducting quantum processors}},
url = {https://doi.org/10.1038/s41467-026-76090-6},
volume = {17},
year = {2026}
}

@article{CarreraVazquez2024,
author = {{Carrera Vazquez}, Almudena and Tornow, Caroline and Rist{\`{e}}, Diego and Woerner, Stefan and Takita, Maika and Egger, Daniel J},
doi = {10.1038/s41586-024-08178-2},
issn = {1476-4687},
journal = {Nature},
number = {8041},
pages = {75--79},
title = {{Combining quantum processors with real-time classical communication}},
url = {https://doi.org/10.1038/s41586-024-08178-2},
volume = {636},
year = {2024}
}

@article{Steinberg2024,
author = {Steinberg, Matthew and Bandi{\'{c}}, Medina and Szkudlarek, Sacha and Almudever, Carmen G and Sarkar, Aritra and Feld, Sebastian},
doi = {10.1038/s41534-024-00909-7},
issn = {2056-6387},
journal = {npj Quantum Information},
number = {1},
pages = {113},
title = {{Lightcone bounds for quantum circuit mapping via uncomplexity}},
url = {https://doi.org/10.1038/s41534-024-00909-7},
volume = {10},
year = {2024}
}

@article{cswp-xy7k,
  title = {Efficient Quantum Simulation for Translationally Invariant Systems},
  author = {Kattem\"olle, Joris and Burkard, Guido},
  journal = {Phys. Rev. Lett.},
  volume = {136},
  issue = {1},
  pages = {010602},
  numpages = {7},
  year = {2026},
  month = {Jan},
  publisher = {American Physical Society},
  doi = {10.1103/cswp-xy7k},
  url = {https://link.aps.org/doi/10.1103/cswp-xy7k}
}

@article{PRXQuantum.4.010313,
  title = {Advantages and Limitations of Quantum Routing},
  author = {Bapat, Aniruddha and Childs, Andrew M. and Gorshkov, Alexey V. and Schoute, Eddie},
  journal = {PRX Quantum},
  volume = {4},
  issue = {1},
  pages = {010313},
  numpages = {20},
  year = {2023},
  month = {Feb},
  publisher = {American Physical Society},
  doi = {10.1103/PRXQuantum.4.010313},
  url = {https://link.aps.org/doi/10.1103/PRXQuantum.4.010313}
}

@article{PhysRevLett.134.020801,
  title = {Long-Range $ZZ$ Interaction via Resonator-Induced Phase in Superconducting Qubits},
  author = {Deng, Xiang and Zheng, Wen and Liao, Xudong and Zhou, Haoyu and Ge, Yangyang and Zhao, Jie and Lan, Dong and Tan, Xinsheng and Zhang, Yu and Li, Shaoxiong and Yu, Yang},
  journal = {Phys. Rev. Lett.},
  volume = {134},
  issue = {2},
  pages = {020801},
  numpages = {7},
  year = {2025},
  month = {Jan},
  publisher = {American Physical Society},
  doi = {10.1103/PhysRevLett.134.020801},
  url = {https://link.aps.org/doi/10.1103/PhysRevLett.134.020801}
}

@article{82cj-lfzy,
  title = {Overhead in Quantum Circuits with Time-Multiplexed Qubit Control},
  author = {Richter, Marvin and Strandberg, Ingrid and Gasparinetti, Simone and Kockum, Anton Frisk},
  journal = {PRX Quantum},
  volume = {7},
  issue = {2},
  pages = {020308},
  numpages = {26},
  year = {2026},
  month = {Apr},
  publisher = {American Physical Society},
  doi = {10.1103/82cj-lfzy},
  url = {https://link.aps.org/doi/10.1103/82cj-lfzy}
}

@article{Eickbusch2025,
author = {Eickbusch, Alec and McEwen, Matt and Sivak, Volodymyr and Bourassa, Alexandre and Atalaya, Juan and Claes, Jahan and Kafri, Dvir and Gidney, Craig and Warren, Christopher W and Gross, Jonathan and Opremcak, Alex and Zobrist, Nicholas and Miao, Kevin C and Roberts, Gabrielle and Satzinger, Kevin J and Bengtsson, Andreas and Neeley, Matthew and Livingston, William P and Greene, Alex and Acharya, Rajeev and {Aghababaie Beni}, Laleh and Aigeldinger, Georg and Alcaraz, Ross and Andersen, Trond I and Ansmann, Markus and Arute, Frank and Arya, Kunal and Asfaw, Abraham and Babbush, Ryan and Ballard, Brian and Bardin, Joseph C and Bilmes, Alexander and Bovaird, Jenna and Bowers, Dylan and Brill, Leon and Broughton, Michael and Browne, David A and Buchea, Brett and Buckley, Bob B and Burger, Tim and Burkett, Brian and Bushnell, Nicholas and Cabrera, Anthony and Campero, Juan and Chang, Hung-Shen and Chiaro, Ben and Chih, Liang-Ying and Cleland, Agnetta Y and Cogan, Josh and Collins, Roberto and Conner, Paul and Courtney, William and Crook, Alexander L and Curtin, Ben and Das, Sayan and {Del Toro Barba}, Alexander and Demura, Sean and {De Lorenzo}, Laura and {Di Paolo}, Agustin and Donohoe, Paul and Drozdov, Ilya K and Dunsworth, Andrew and Elbag, Aviv Moshe and Elzouka, Mahmoud and Erickson, Catherine and Ferreira, Vinicius S and {Flores Burgos}, Leslie and Forati, Ebrahim and Fowler, Austin G and Foxen, Brooks and Ganjam, Suhas and Garcia, Gonzalo and Gasca, Robert and Genois, {\'{E}}lie and Giang, William and Gilboa, Dar and Gosula, Raja and {Grajales Dau}, Alejandro and Graumann, Dietrich and Ha, Tan and Habegger, Steve and Hamilton, Michael C and Hansen, Monica and Harrigan, Matthew P and Harrington, Sean D and Heslin, Stephen and Heu, Paula and Higgott, Oscar and Hiltermann, Reno and Hilton, Jeremy and Huang, Hsin-Yuan and Huff, Ashley and Huggins, William J and Jeffrey, Evan and Jiang, Zhang and Jin, Xiaoxuan and Jones, Cody and Joshi, Chaitali and Juhas, Pavol and Kabel, Andreas and Kang, Hui and Karamlou, Amir H and Kechedzhi, Kostyantyn and Khaire, Trupti and Khattar, Tanuj and Khezri, Mostafa and Kim, Seon and Kobrin, Bryce and Korotkov, Alexander N and Kostritsa, Fedor and Kreikebaum, John Mark and Kurilovich, Vladislav D and Landhuis, David and Lange-Dei, Tiano and Langley, Brandon W and Lau, Kim-Ming and Ledford, Justin and Lee, Kenny and Lester, Brian J and {Le Guevel}, Lo{\"{i}}ck and Li, Wing Yan and Lill, Alexander T and Locharla, Aditya and Lucero, Erik and Lundahl, Daniel and Lunt, Aaron and Madhuk, Sid and Maloney, Ashley and Mandr{\`{a}}, Salvatore and Martin, Leigh S and Martin, Orion and Maxfield, Cameron and McClean, Jarrod R and Meeks, Seneca and Megrant, Anthony and Molavi, Reza and Molina, Sebastian and Montazeri, Shirin and Movassagh, Ramis and Newman, Michael and Nguyen, Anthony and Nguyen, Murray and Ni, Chia-Hung and Oas, Logan and Orosco, Raymond and Ottosson, Kristoffer and Pizzuto, Alex and Potter, Rebecca and Pritchard, Orion and Quintana, Chris and Ramachandran, Ganesh and Reagor, Matthew J and Rhodes, David M and Rosenberg, Eliott and Rossi, Elizabeth and Sankaragomathi, Kannan and Schurkus, Henry F and Shearn, Michael J and Shorter, Aaron and Shutty, Noah and Shvarts, Vladimir and Small, Spencer and Smith, W Clarke and Springer, Sofia and Sterling, George and Suchard, Jordan and Szasz, Aaron and Sztein, Alex and Thor, Douglas and Tomita, Eifu and Torres, Alfredo and Torunbalci, M Mert and Vaishnav, Abeer and Vargas, Justin and Vdovichev, Sergey and Vidal, Guifre and {Vollgraff Heidweiller}, Catherine and Waltman, Steven and Waltz, Jonathan and Wang, Shannon X and Ware, Brayden and Weidel, Travis and White, Theodore and Wong, Kristi and Woo, Bryan W K and Woodson, Maddy and Xing, Cheng and Yao, Z Jamie and Yeh, Ping and Ying, Bicheng and Yoo, Juhwan and Yosri, Noureldin and Young, Grayson and Zalcman, Adam and Zhang, Yaxing and Zhu, Ningfeng and Boixo, Sergio and Kelly, Julian and Smelyanskiy, Vadim and Neven, Hartmut and Bacon, Dave and Chen, Zijun and Klimov, Paul V and Roushan, Pedram and Neill, Charles and Chen, Yu and Morvan, Alexis},
doi = {10.1038/s41567-025-03070-w},
issn = {1745-2481},
journal = {Nature Physics},
number = {12},
pages = {1994--2001},
title = {{Demonstration of dynamic surface codes}},
url = {https://doi.org/10.1038/s41567-025-03070-w},
volume = {21},
year = {2025}
}

@article{
doi:10.1126/science.adz8659,
author = {David D. Awschalom  and Hannes Bernien  and Ronald Hanson  and William D. Oliver  and Jelena Vučković },
title = {Challenges and opportunities for quantum information hardware},
journal = {Science},
volume = {390},
number = {6777},
pages = {1004-1010},
year = {2025},
doi = {10.1126/science.adz8659},
URL = {https://www.science.org/doi/abs/10.1126/science.adz8659}}

@article{Zhou2023,
author = {Zhou, Chao and Lu, Pinlei and Praquin, Matthieu and Chien, Tzu-Chiao and Kaufman, Ryan and Cao, Xi and Xia, Mingkang and Mong, Roger S K and Pfaff, Wolfgang and Pekker, David and Hatridge, Michael},
doi = {10.1038/s41534-023-00723-7},
issn = {2056-6387},
journal = {npj Quantum Information},
number = {1},
pages = {54},
title = {{Realizing all-to-all couplings among detachable quantum modules using a microwave quantum state router}},
url = {https://doi.org/10.1038/s41534-023-00723-7},
volume = {9},
year = {2023}
}

@article{PhysRevX.14.041030,
  title = {Modular Quantum Processor with an All-to-All Reconfigurable Router},
  author = {Wu, Xuntao and Yan, Haoxiong and Andersson, Gustav and Anferov, Alexander and Chou, Ming-Han and Conner, Christopher R. and Grebel, Joel and Joshi, Yash J. and Li, Shiheng and Miller, Jacob M. and Povey, Rhys G. and Qiao, Hong and Cleland, Andrew N.},
  journal = {Phys. Rev. X},
  volume = {14},
  issue = {4},
  pages = {041030},
  numpages = {22},
  year = {2024},
  month = {Nov},
  publisher = {American Physical Society},
  doi = {10.1103/PhysRevX.14.041030},
  url = {https://link.aps.org/doi/10.1103/PhysRevX.14.041030}
}

@misc{wu2026efficientnqubitentanglingoperations,
      title={Efficient $n$-qubit entangling operations via a superconducting quantum router}, 
      author={Xuntao Wu and Haoxiong Yan and Gustav Andersson and Alexander Anferov and Christopher R. Conner and Yash J. Joshi and Bayan Karimi and Amber M. King and Shiheng Li and Howard L. Malc and Jacob M. Miller and Harsh Mishra and Hong Qiao and Minseok Ryu and Jian Shi and Andrew N. Cleland},
      year={2026},
      eprint={2604.15432},
      archivePrefix={arXiv},
      primaryClass={quant-ph},
      url={https://arxiv.org/abs/2604.15432}, 
}

@article{c661-yr2z,
  title = {Symmetrically Threaded Superconducting Quantum Interference Devices as Next-Generation Kerr-Cat Qubits},
  author = {Bhandari, Bibek and Huang, Irwin and Hajr, Ahmed and Yanik, Kagan and Qing, Bingcheng and Wang, Ke and Santiago, David I. and Dressel, Justin and Siddiqi, Irfan and Jordan, Andrew N.},
  journal = {PRX Quantum},
  volume = {6},
  issue = {3},
  pages = {030338},
  numpages = {30},
  year = {2025},
  month = {Aug},
  publisher = {American Physical Society},
  doi = {10.1103/c661-yr2z},
  url = {https://link.aps.org/doi/10.1103/c661-yr2z}
}

@article{PhysRevX.14.031040,
  title = {Observation of Pairwise Level Degeneracies and the Quantum Regime of the Arrhenius Law in a Double-Well Parametric Oscillator},
  author = {Frattini, Nicholas E. and Corti\~nas, Rodrigo G. and Venkatraman, Jayameenakshi and Xiao, Xu and Su, Qile and Lei, Chan U. and Chapman, Benjamin J. and Joshi, Vidul R. and Girvin, S. M. and Schoelkopf, Robert J. and Puri, Shruti and Devoret, Michel H.},
  journal = {Phys. Rev. X},
  volume = {14},
  issue = {3},
  pages = {031040},
  numpages = {28},
  year = {2024},
  month = {Sep},
  publisher = {American Physical Society},
  doi = {10.1103/PhysRevX.14.031040},
  url = {https://link.aps.org/doi/10.1103/PhysRevX.14.031040}
}

@article{10.1063/1.4984142,
    author = {Frattini, N. E. and Vool, U. and Shankar, S. and Narla, A. and Sliwa, K. M. and Devoret, M. H.},
    title = {3-wave mixing Josephson dipole element},
    journal = {Applied Physics Letters},
    volume = {110},
    number = {22},
    pages = {222603},
    year = {2017},
    month = {05},
    issn = {0003-6951},
    doi = {10.1063/1.4984142},
    url = {https://doi.org/10.1063/1.4984142}
}

@article{PhysRevLett.133.150602,
  title = {Quantum Fourier Transform Using Dynamic Circuits},
  author = {B\"aumer, Elisa and Tripathi, Vinay and Seif, Alireza and Lidar, Daniel and Wang, Derek S.},
  journal = {Phys. Rev. Lett.},
  volume = {133},
  issue = {15},
  pages = {150602},
  numpages = {7},
  year = {2024},
  month = {Oct},
  publisher = {American Physical Society},
  doi = {10.1103/PhysRevLett.133.150602},
  url = {https://link.aps.org/doi/10.1103/PhysRevLett.133.150602}
}

@article{
doi:10.1126/sciadv.adm6761,
author = {Ruslan Shaydulin  and Changhao Li  and Shouvanik Chakrabarti  and Matthew DeCross  and Dylan Herman  and Niraj Kumar  and Jeffrey Larson  and Danylo Lykov  and Pierre Minssen  and Yue Sun  and Yuri Alexeev  and Joan M. Dreiling  and John P. Gaebler  and Thomas M. Gatterman  and Justin A. Gerber  and Kevin Gilmore  and Dan Gresh  and Nathan Hewitt  and Chandler V. Horst  and Shaohan Hu  and Jacob Johansen  and Mitchell Matheny  and Tanner Mengle  and Michael Mills  and Steven A. Moses  and Brian Neyenhuis  and Peter Siegfried  and Romina Yalovetzky  and Marco Pistoia },
title = {Evidence of scaling advantage for the quantum approximate optimization algorithm on a classically intractable problem},
journal = {Science Advances},
volume = {10},
number = {22},
pages = {eadm6761},
year = {2024},
doi = {10.1126/sciadv.adm6761},
URL = {https://www.science.org/doi/abs/10.1126/sciadv.adm6761}}

@inproceedings{10.1145/3297858.3304023,
author = {Li, Gushu and Ding, Yufei and Xie, Yuan},
title = {Tackling the Qubit Mapping Problem for NISQ-Era Quantum Devices},
year = {2019},
isbn = {9781450362405},
publisher = {Association for Computing Machinery},
address = {New York, NY, USA},
doi = {10.1145/3297858.3304023},
booktitle = {Proceedings of the Twenty-Fourth International Conference on Architectural Support for Programming Languages and Operating Systems},
pages = {1001–1014},
numpages = {14},
location = {Providence, RI, USA},
series = {ASPLOS 19}
}

@ARTICLE{11267428,
  author={Kondoth, Lashmi and Shankaran, Rajan and Sheng, Quan Z. and Kuantama, Endrowednes and Ni, Wei},
  journal={IEEE Communications Surveys \& Tutorials}, 
  title={A Survey of Adaptive Routing Techniques in Two-Dimensional Network-on-Chip Architectures}, 
  year={2026},
  volume={28},
  number={},
  pages={3195-3234},
  doi={10.1109/COMST.2025.3636503}}

@ARTICLE{52203,
  author={Tobagi, F.A.},
  journal={Proceedings of the IEEE}, 
  title={Fast packet switch architectures for broadband integrated services digital networks}, 
  year={1990},
  volume={78},
  number={1},
  pages={133-167},
  doi={10.1109/5.52203}}

@article{PhysRevA.69.062320,
  title = {Cavity quantum electrodynamics for superconducting electrical circuits: An architecture for quantum computation},
  author = {Blais, Alexandre and Huang, Ren-Shou and Wallraff, Andreas and Girvin, S. M. and Schoelkopf, R. J.},
  journal = {Phys. Rev. A},
  volume = {69},
  issue = {6},
  pages = {062320},
  numpages = {14},
  year = {2004},
  month = {Jun},
  publisher = {American Physical Society},
  doi = {10.1103/PhysRevA.69.062320},
  url = {https://link.aps.org/doi/10.1103/PhysRevA.69.062320}
}

@article{cqed,
Author = {Blais, Alexandre and Grimsmo, Arne L. and Girvin, S. M. and Wallraffe,
   Andreas},
Title = {Circuit quantum electrodynamics},
Journal = {REVIEWS OF MODERN PHYSICS},
Year = {2021},
Volume = {93},
Number = {2},
Month = {MAY 19},
DOI = {10.1103/RevModPhys.93.025005},
Article-Number = {025005},
ISSN = {0034-6861},
EISSN = {1539-0756},
ResearcherID-Numbers = {Grimsmo, Arne/AAM-3155-2020
   Girvin, Steven/C-1471-2012
   Wallraff, Andreas/C-2130-2009
   Blais, Alexandre/E-3315-2010},
ORCID-Numbers = {Wallraff, Andreas/0000-0002-3476-4485
   Blais, Alexandre/0000-0002-0591-2012},
Unique-ID = {WOS:000655978700001},
}

@article{KrantzReview,
Author = {Krantz, P. and Kjaergaard, M. and Yan, F. and Orlando, T. P. and
   Gustavsson, S. and Oliver, W. D.},
Title = {A quantum engineer's guide to superconducting qubits},
Journal = {APPLIED PHYSICS REVIEWS},
Year = {2019},
Volume = {6},
Number = {2},
Month = {JUN},
DOI = {10.1063/1.5089550},
Article-Number = {021318},
ISSN = {1931-9401},
ResearcherID-Numbers = {Yan, Fei/HLX-7286-2023
   Kjaergaard, Morten/JFK-1649-2023
   Gustavsson, Simon/B-5391-2012
   Krantz, Philip/Q-3701-2016},
ORCID-Numbers = {Kjaergaard, Morten/0000-0002-1179-581X
   Gustavsson, Simon/0000-0002-7069-1025
   Oliver, William/0000-0001-8041-0824
   Krantz, Philip/0000-0002-8553-3353},
Unique-ID = {WOS:000474435200017},
}

@article{JOHANSSON20121760,
title = {QuTiP: An open-source Python framework for the dynamics of open quantum systems},
journal = {Computer Physics Communications},
volume = {183},
number = {8},
pages = {1760-1772},
year = {2012},
issn = {0010-4655},
doi = {https://doi.org/10.1016/j.cpc.2012.02.021},
url = {https://www.sciencedirect.com/science/article/pii/S0010465512000835},
author = {J.R. Johansson and P.D. Nation and Franco Nori}
}

@article{PhysRevLett.129.150504,
  title = {Universal Fidelity Reduction of Quantum Operations from Weak Dissipation},
  author = {Abad, Tahereh and Fern\'andez-Pend\'as, Jorge and Frisk Kockum, Anton and Johansson, G\"oran},
  journal = {Phys. Rev. Lett.},
  volume = {129},
  issue = {15},
  pages = {150504},
  numpages = {6},
  year = {2022},
  month = {Oct},
  publisher = {American Physical Society},
  doi = {10.1103/PhysRevLett.129.150504},
  url = {https://link.aps.org/doi/10.1103/PhysRevLett.129.150504}
}

@article{Abad2025impactofdecoherence,
  doi = {10.22331/q-2025-04-03-1684},
  url = {https://doi.org/10.22331/q-2025-04-03-1684},
  title = {Impact of decoherence on the fidelity of quantum gates leaving the computational subspace},
  author = {Abad, Tahereh and Schattner, Yoni and Kockum, Anton Frisk and Johansson, G{\"{o}}ran},
  journal = {{Quantum}},
  issn = {2521-327X},
  publisher = {{Verein zur F{\"{o}}rderung des Open Access Publizierens in den Quantenwissenschaften}},
  volume = {9},
  pages = {1684},
  month = apr,
  year = {2025}
}

@article{BRAVYI20112793,
title = {Schrieffer–Wolff transformation for quantum many-body systems},
journal = {Annals of Physics},
volume = {326},
number = {10},
pages = {2793-2826},
year = {2011},
issn = {0003-4916},
doi = {https://doi.org/10.1016/j.aop.2011.06.004},
url = {https://www.sciencedirect.com/science/article/pii/S0003491611001059},
author = {Sergey Bravyi and David P. DiVincenzo and Daniel Loss}
}

@article{PhysRevApplied.10.054062,
  title = {Tunable Coupling Scheme for Implementing High-Fidelity Two-Qubit Gates},
  author = {Yan, Fei and Krantz, Philip and Sung, Youngkyu and Kjaergaard, Morten and Campbell, Daniel L. and Orlando, Terry P. and Gustavsson, Simon and Oliver, William D.},
  journal = {Phys. Rev. Appl.},
  volume = {10},
  issue = {5},
  pages = {054062},
  numpages = {9},
  year = {2018},
  month = {Nov},
  publisher = {American Physical Society},
  doi = {10.1103/PhysRevApplied.10.054062},
  url = {https://link.aps.org/doi/10.1103/PhysRevApplied.10.054062}
}

@article{PhysRevA.87.022309,
  title = {High-fidelity controlled-${\ensuremath{\sigma}}^{Z}$ gate for resonator-based superconducting quantum computers},
  author = {Ghosh, Joydip and Galiautdinov, Andrei and Zhou, Zhongyuan and Korotkov, Alexander N. and Martinis, John M. and Geller, Michael R.},
  journal = {Phys. Rev. A},
  volume = {87},
  issue = {2},
  pages = {022309},
  numpages = {19},
  year = {2013},
  month = {Feb},
  publisher = {American Physical Society},
  doi = {10.1103/PhysRevA.87.022309},
  url = {https://link.aps.org/doi/10.1103/PhysRevA.87.022309}
}

@article{PhysRevLett.125.200503,
  title = {High-Contrast $ZZ$ Interaction Using Superconducting Qubits with Opposite-Sign Anharmonicity},
  author = {Zhao, Peng and Xu, Peng and Lan, Dong and Chu, Ji and Tan, Xinsheng and Yu, Haifeng and Yu, Yang},
  journal = {Phys. Rev. Lett.},
  volume = {125},
  issue = {20},
  pages = {200503},
  numpages = {6},
  year = {2020},
  month = {Nov},
  publisher = {American Physical Society},
  doi = {10.1103/PhysRevLett.125.200503},
  url = {https://link.aps.org/doi/10.1103/PhysRevLett.125.200503}
}

\end{document}